\documentclass[aps,amssymb,noshowpacs,superscriptaddress,nofootinbib]{revtex4} \usepackage{mathrsfs} \usepackage{graphicx} \usepackage{enumerate}

\begin{document}

\title{Connection dynamics of a gauge theory of gravity coupled with matter}

\author{Jian Yang} \email{yjccnuphy@yahoo.com.cn} \affiliation{School of Science, Beijing University of Posts and Telecommunications,\\ Beijing 100876, China.}
\author{Kinjal
Banerjee} \email{kinjalb@gmail.com} \affiliation{Department of Physics, Beijing Normal University, \\Beijing 100875, China.} \affiliation{BITS Pilani, K.K. Birla Goa Campus, NH 17B Zuarinagar, Goa 403726, India.}
\author{Yongge Ma\footnote{Corresponding author}}
\email{mayg@bnu.edu.cn} \affiliation{Department of Physics, Beijing Normal University, \\Beijing 100875, China.}

\date{\today}
\begin{abstract}

We study the coupling of the gravitational action, which is a linear combination of the Hilbert-Palatini term and the quadratic torsion term, to the action of Dirac fermions. The system possesses local Poincare invariance and hence belongs to Poincare gauge theory with
matter. The complete Hamiltonian analysis of the theory is carried out without gauge fixing but under certain ansatz on the coupling parameters, which leads to a consistent connection dynamics with second-class constraints and torsion. After performing a partial gauge fixing, all second-class constraints can be solved,
and a $SU(2)$-connection dynamical formalism of the theory can be obtained. Hence, the techniques of loop quantum gravity can be employed to quantize this Poincare gauge theory with non-zero torsion. Moreover, the Barbero-Immirzi parameter in loop quantum gravity
acquires its physical meaning as the coupling parameter between the Hilbert-Palatini term and the quadratic torsion term in this gauge theory of gravity.\\ { }
\end{abstract}

\maketitle


\section{Introduction}

General Relativity(GR) has been very successful in describing universe at large scales. However, it is believed that we have to develop a quantum theory of gravity for a consistent description of nature. One of the reasons that classical GR cannot be consistent can be
seen from the Einstein's equations which relate gravitational and matter degrees of freedom. While the gravitational part is classical and is encoded in the Einstein tensor, since matter interactions are very well described by quantum field theory, we need to use some
quantum version of the stress energy tensor for the matter part. This would imply that a consistent coupling of matter and gravity for all energy scales requires both of them to be quantized.

Einstein's equations can be obtained via an action principle starting from the first-order Hilbert-Palatini action.  However, if we consider fermionic matter sources, the equations of motion from this action will not provide the torsion-free condition of vacuum case.
Hence, we have to either allow for torsion or make some suitable modification of the action. (See \cite{hehl} and references therein for a comprehensive account of torsion in gravity). So, if one wants to start with first-order action, it is very possible that quantum
theory of gravity would incorporate torsion in its formalism in order to consistently couple gravity to fermions. Among various attempts to look for a quantum gravity theory, gauge theories of gravity are very attractive since the idea of gauge invariance has already
been successful in the description of other fundamental interactions. Local gauge invariance is a key concept in Yang-Mills theory. Together with Poincare symmetry, it lays the foundation of standard model in particle physics. Localization of Poincare symmetry leads to
Poincare Gauge Theory(PGT) of gravity. One of the key features in PGT is that, in general, gravity is not only represented as curvature but also as torsion of space-time. GR is a special case of PGT when torsion equals zero.

PGT provides a very convenient framework for studying theories with torsion.  A number of actions which satisfy local Poincare symmetry have been analyzed by various researchers (Refs.\cite{PGT,PGT2} provide the comprehensive review and bibliography of the progress made
in PGT). However, one of the drawbacks of PGT is that its Hamiltonian formulation is usually very complicated. Although Hamiltonian analysis is performed for many models in PGT, the results are at a formal level without explicit expressions of the additional required
second-class constraints. From the point of view of canonical quantization, it is essential to have a well-defined consistent Hamiltonian theory at the classical level. Such an ingredient is missing if we want to incorporate torsion into candidate quantum gravity models
constructed from PGT. Moreover, the internal gauge group in PGT is in general non-compact, while most of the standard tools developed in quantum field theory apply to gauge theories with compact gauge groups.

There exists a well-known $SU(2)$ gauge theory formulation of canonical GR \cite{ashtekar,barbimm}, where the basic variables are the densitized triad and Ashtekar-Barbero connection. A candidate canonical quantum gravity theory known as Loop Quantum Gravity (LQG)
\cite{thiemannbook,rovellibook,alrev,marev} can be constructed starting from the connection dynamical formulation. Moreover, LQG can also be extended to some modified gravity theories such as, $f(R)$ theories \cite{ZM11,ZM11b} and scalar-tensor theories \cite{ZM11c}.
However, the action of GR from which the connection dynamics can be derived is not the standard Hilbert-Palatini action. An additional term known as the Holst term has to be added to the standard Hilbert-Palatini action in order to rewrite GR as a $SU(2)$ gauge theory
\cite{holst,sa}. It is customary to multiply the additional Holst term with a coupling constant $\gamma$ known as the Barbero-Immirzi parameter. Classically these two actions are equivalent in vacuum case, since the additional Holst term does not affect the equations of motion although
it is not a total derivative. The parameter $\gamma$ does not appear in the classical equations of motion. This is because the Holst term differs from a total derivative known as the Nieh-Yan term \cite{niehyan} by a term quadratic in torsion (for the exact relations
between them see \cite{shyam,kinjal}).
 Since the torsion term is zero when there is no fermionic matter, the
 Nieh-Yan term and the Holst term are same, and hence the connection
dynamics obtained from adding either term to the Hilbert-Palatini action would be equivalent. It has been shown  that a $SU(2)$ gauge theory can also be constructed from an action containing  the standard Hilbert-Palatini term  and the Nieh-Yan term \cite{shyam}.
However, when there are fermions, the $T^2$ term is not zero and the the difference in the Holst term and the Nieh-Yan term shows up. In Ref.\cite{perezrovelli} it was found that adding the standard fermion action along with the Holst term leads to equations of motion
which depend on $\gamma$ and are therefore not equivalent to standard GR with fermions. The difference arises because the Holst term is not a total derivative. In Ref.\cite{shyam} it was shown that there is no such issue if the full Nieh-Yan term is used. An alternative
possibility of modifying the fermion action to be \textit{non-minimally coupled} has been analyzed in detail in Refs.\cite{mercuri,martinfermion} and also in Refs.\cite{alexandrov,friedel}. The additional piece in fermion action cancels the contribution of the Holst
piece if the coupling constants are chosen accordingly (see \cite{shyamreview} for a recent account of these issues). In the absence of direct experimental or observational evidence of quantum gravity and of torsion, it is not clear which action should be the
appropriate starting point for quantization, particularly from the perspective of LQG. It is therefore very important to study all the different possibilities. However to apply the LQG techniques, it is essential to first reformulate these candidates as gauge theories
with a compact gauge group.

In this series of works, instead of the Holst piece of the Nieh-Yan term, we consider the $T^2$ piece. In Ref.\cite{1stpaper} we considered the vacuum case, i.e. an action with only this $T^2$ term along with the standard Hilbert-Palatini term. An arbitrary coupling
constant $\alpha$ between the Hilbert-Palatini and $T^2$ terms was employed. There it was shown that, although we started from an action with explicit torsion dependence, the constraint equations imply that torsion is zero, and hence we go back to standard GR. This is
consistent with the results that there is no torsion in the absence of spinors. The variables we choose are motivated by PGT. But unlike
other analysis in PGT we obtain explicit expressions of the second-class constraints.

In this paper, we add Dirac fermions to the action and apply the techniques developed in Ref.\cite{1stpaper} to carry out the Hamiltonian analysis. We consider the fermions to be non-minimally coupled, because the $T^2$ term is not a total derivative and indeed, by
proper choice of the two coefficients, the contribution of the additional non-minimal piece is canceled by the contribution of the torsion piece. Also the relation between torsion and the fermions we obtain is the same as the one obtained in Ref.\cite{shyam} with
Nieh-Yan term and minimally coupled fermion action. To the best of our knowledge, this is the first action with explicit torsion terms which has been reformulated as a Hamiltonian $SU(2)$ gauge theory. The new connection we obtain is algebraically same as the standard
Ashtekar-Barbero connection but is valid even in the presence of explicit torsion dependent terms of the form we have chosen. This is unlike the standard derivation of the Ashtekar-Barbero formalism\cite{barbimm} which was done for the torsion-free case. The coupling
parameter $\alpha$ in our action plays the role of Barbero-Immirzi parameter. The classical system we obtain in this paper can subsequently be loop quantized using the tools already developed in LQG. Also, Hamiltonian formulation of theories with torsion are usually very complicated. We think that the techniques developed in this and the previous paper \cite{1stpaper} can be used for analyzing other similar actions with torsion terms. If that is
possible, then the general programme of loop quantization can be applied to a much wider class of theories which include torsion.

The paper is organized as follows. In section \ref{sec2} we give the explicit expression of the action with which we start and derive the equations of motion for the coupled system. It is shown that under certain ansatz on the coupling parameters, the dynamical system we obtain is equivalent to the standard Palatini formulation of GR minimally coupled to
fermions. In section \ref{sec3} we perform a $3+1$ decomposition of this action and perform the Hamiltonian
analysis under the ansatz but without fixing time gauge. Having obtained a consistent Hamiltonian system, we fix time gauge and then solve the second class constraints in section \ref{sec4}. Fixing the time gauge also breaks the $SO(1,3)$ gauge invariance to $SU(2)$. Then in section
\ref{sec5} a new connection which is conjugate to the densitized triad is derived, and thus we obtain a $SU(2)$ gauge theory. Our analysis has several novel and peculiar features. We conclude with a discussion of these and some comparison of our results with those
obtained by using the Holst and Nieh-Yan terms in section \ref{sec6}. We will restrict ourselves to 4 dimensions. The Greek letters {$ \mu, \nu \dots$} refer to space-time indices while the uppercase Latin letters {$ I,J \dots$} refer to the internal $SO(1,3)$ indices.
Our spacetime metric signature is $(-+++)$. Later when we do the $3+1$ decomposition of spacetime, we will use the lowercase Latin letters from the beginning of the alphabet {$ a,b,\dots$} to represent the spatial indices. After we reduce the symmetry group to $SU(2)$,
the internal indices will be represented by lowercase Latin letters from the middle of the alphabet {$ i,j \dots$}.


\section{The Action}\label{sec2}

In this paper we consider an action which has three pieces, a Hilbert-Palatini term, a term quadratic in torsion and a term for the massless fermionic matter. It reads
\begin{eqnarray}
S = \int d^{4}x \mathcal{L}=S_{HP} + \alpha S_T  + S_M ~~,~~ \hspace{6em}
\label{action}
\end{eqnarray}
where
\begin{eqnarray}
 && \hspace{3em} S_{HP} =\int {\mbox d}^4 x ~e R= \int {\mbox d}^4 x e e^{\mu}_I e^{\nu}_J R_{\mu \nu}^{~~ IJ}(\omega_\mu^{~IJ}) ~~,~~ \nonumber\\
&& \hspace{3em} S_T  = \frac{1}{8}\int {\mbox d}^4 x \epsilon^{\mu \nu \rho \sigma} T^{I}_{~ \mu \nu} T_{I \rho \sigma } ~~,~~ \nonumber\\ && \hspace{3em} S_M = i  \int {\mbox d}^4 x e \left[ \overline{\lambda}(1 + i \varepsilon \gamma_5) \gamma^\mu D_\mu \lambda -
\overline{D_\mu \lambda}\gamma^\mu (1 + i \varepsilon \gamma_5) \lambda \right] ~~.~~  \nonumber
\end{eqnarray}
Here $e^{\mu}_I$ is the tetrad, $e$ denotes the absolute value of the determinant of the co-tetrad, $\omega_\mu^{~IJ}$ is the spacetime spin-connection which
is not torsion-free, $\epsilon^{\mu \nu \rho \sigma}$ denotes the 4-dimensional Levi-Civita tensor density, and the covariant derivatives in the fermion action read,
 \begin{eqnarray}
 D_\mu \lambda = \partial_\mu \lambda +\frac{1}{2} ~\omega_\mu^{~IJ}
\sigma_{IJ}~\lambda ~~~;~~~ \overline{D_\mu \lambda} = \partial_\mu \overline{\lambda} - \frac{1}{2} ~\overline{\lambda} ~\omega_\mu^{~IJ} \sigma_{IJ}. \nonumber
\end{eqnarray}
 Note that we denote $\gamma^\mu = \gamma^I e^\mu_I$ with 4-dimensional Dirac matrices
$\gamma^I$ , $\sigma_{IJ} := \frac{1}{4}[\gamma_I,\gamma_J]$ and $\gamma_5 := i \gamma_0 \gamma_1 \gamma_2 \gamma_3$. Our conventions regarding the Dirac matrices and their properties are given in Appendix (\ref{app1}). Note also that $\lambda$ and  $\overline{\lambda}
:= \lambda^\dagger \gamma^0$, representing the fermionic degrees of freedom, are 4-dimensional row and column vector respectively. Further,
 \begin{eqnarray}
 R_{\mu \nu}^{~~ IJ} &=& \partial_{[\mu} \omega_{\nu]}^{~IJ} + \omega_{[\mu}^{~IK} \omega_{\nu] K}^{~~~J} , \\
T^{I}_{~ \mu \nu}  &=& \partial_{[\mu} e_{\nu]}^{I} +  \omega_{[\mu~|J|}^{~~I}e_{\nu]}^{J}
\end{eqnarray}
are the definitions for curvature and torsion respectively\footnote{Our conventions of symmetrization and antisymmetrization are $A^{(ab)}:=A^{ab} + A^{ba}$ and
$A^{[ab]}:=A^{ab} - A^{ba}$ respectively}. It should be noted that the boundary terms of the action (\ref{action}) are neglected. This means that we either consider a compact spacetime without boundary or assume suitable boundary conditions for the fields configuration
such that there is no boundary term. It is obvious that this action is invariant under local Poincare transformations \cite{1stpaper}. We will be working in the first-order formalism and hence both the co-tetrad $e_{\mu}^I$ and the spin connection $\omega_{\mu}^{~IJ}$
are treated as independent fields. Our covariant derivative $D_\mu$ acts in the following way:
\begin{eqnarray}
D_\mu e_\nu^I:= \partial_\mu e_\nu^I +  \omega_{\mu ~ J}^{~I} e_\nu^J.  \nonumber
\end{eqnarray}
Note that the coupling parameter $\alpha$ in action (\ref{action}) is a non-zero real number.
The parameter $\varepsilon$ in the matter action denotes nonminimal coupling and with $\varepsilon=0$ we get back minimally coupled Fermion action.

Let us consider the Lagrangian equations of motion. The variations of action (\ref{action}) yield
\begin{eqnarray}
\frac{\delta S}{\delta\omega_{\mu}^{~IJ}}&=&\frac{1}{2}\epsilon^{\mu \nu \rho \sigma}  e_{\nu}^K D_{[\rho}e_{\sigma]}^L \left[\frac{\alpha}{2} (\eta_{JK} \eta_{IL} -\eta_{IK} \eta_{JL}) - \epsilon_{IJKL}\right] \nonumber\\ && - \frac{1}{2} e e_K^{\mu} \overline{\lambda}
\gamma_5 \gamma_L \lambda \left[\varepsilon(\eta_{IK} \eta_{JL} -\eta_{JK} \eta_{IL}) + \epsilon_{IJKL}\right] = 0, \label{omega1}\\
\frac{\delta S}{\delta e^{K}_{\alpha}}&=&e e^{\alpha}_K e^{\mu}_I e^{\nu}_J R_{\mu \nu}^{~~ IJ} - 2 e  e^{\alpha}_I  e^{\mu}_K e^{\nu}_J
R_{\mu \nu}^{~~ IJ} + \frac{\alpha}{2} \left(D_{\beta}[\epsilon^{\alpha \beta \gamma \delta} D_{[\gamma}e_{\delta]K}]\right) \nonumber\\
&& +iee^{\alpha}_{[K}e^{\mu}_{I]}[\overline{\lambda}(1+i\varepsilon\gamma_{5})\gamma^{I}D_{\mu}\lambda
-\overline{D_{\mu}\lambda}\gamma^{I}(1+i\varepsilon\gamma_{5})\lambda]=0 ,\label{tetrad1}\\
\frac{\delta S}{\delta \lambda}&=&i[-D_{\mu}(e\overline{\lambda}(1+i\varepsilon\gamma_{5})\gamma^{I}e_{I}^{\mu})-
e\overline{D_{\mu}\lambda} \gamma^{I}e_{I}^{\mu}(1+i\varepsilon\gamma_{5})]=0,\label{lambda1}\\
\frac{\delta S}{\delta \overline{\lambda}}&=&i[e(1+i\varepsilon\gamma_{5})\gamma^{I}e_{I}^{\mu}D_{\mu}\lambda+
D_{\mu}(e\gamma^{I}e_{I}^{\mu} (1+i\varepsilon\gamma_{5})\lambda)]=0. \label{lambdabar1}
\end{eqnarray}
The parameter $\varepsilon$, in general, has no relation with the parameter $\alpha$. However if we choose the ansatz
$\varepsilon = \frac{\alpha}{2}$ the equations of motion would be simplified. Let us consider the equations of motion of the spin connection.
If we choose $\varepsilon=\frac{\alpha}{2}$, Eq. (\ref{omega1}) is reduced to
\begin{eqnarray}
 \frac{\delta S}{\delta\omega_{\mu}^{~IJ}}&=&(\frac{1}{2}\epsilon^{\mu \nu \rho \sigma}  e_{\nu}^K
D_{[\rho}e_{\sigma]}^L+\frac{1}{2} e e^{\mu K} \overline{\lambda} \gamma_5 \gamma^L \lambda) \left[\frac{\alpha}{2} (\eta_{JK} \eta_{IL} -
\eta_{IK} \eta_{JL}) - \epsilon_{IJKL}\right] = 0. \label{omega2}
\end{eqnarray}
 Denoting
 \begin{equation}
\frac{1}{2}\epsilon^{\mu \nu \rho \sigma}  e_{\nu}^K D_{[\rho}e_{\sigma]}^L+\frac{1}{2} e e^{\mu K} \overline{\lambda} \gamma_5 \gamma^L \lambda=
s^{\mu KL},
\end{equation}
 Eq. (\ref{omega2}) implies $s^{\mu [KL]}=0$, which is $\alpha$-independent.
Hence the $\alpha$ term in Eq.(\ref{omega2}) will disappear from the equations of motion of $\omega_{\mu}^{~IJ}$. Using this result and the
identity $\epsilon^{\mu\rho\nu\sigma}\epsilon_{IJKL}e^{K}_{\nu}e^{L}_{\sigma}=2ee^{\mu}_{[I}e^{\rho}_{J]}$, it can  be shown, after some
calculation, that for the case $\varepsilon=\frac{\alpha}{2}$, the equations of motion of $e^{K}_{\alpha}$ reduce to the standard form given by
\begin{eqnarray}
\frac{\delta S}{\delta e^{K}_{\alpha}}&=&e e^{\alpha}_K e^{\mu}_I e^{\nu}_J R_{\mu \nu}^{~~ IJ} - 2 e  e^{\alpha}_I  e^{\mu}_K e^{\nu}_J
R_{\mu \nu}^{~~ IJ}+iee^{\alpha}_{[K}e^{\mu}_{I]}(\overline{\lambda}\gamma^{I}D_{\mu}\lambda
-\overline{D_{\mu}\lambda}\gamma^{I}\lambda)=0. \label{tetrad2}
\end{eqnarray}
Further, using the fact that $D_{\mu}(ee^{\mu}_{I}) = 0$, it can be easily shown that the $\varepsilon$ dependence drops out from the equations
of motion of the fermion degrees of freedom $\lambda$ and $\overline{\lambda}$ \cite{martinfermion}, leaving
\begin{eqnarray}
\frac{\delta S}{\delta \lambda}&=&i[-D_{\mu}(e\overline{\lambda}\gamma^{I}e_{I}^{\mu})-
e\overline{D_{\mu}\lambda} \gamma^{I}e_{I}^{\mu}]=0,\label{lambda2}\\
\frac{\delta S}{\delta \overline{\lambda}}&=&i[e\gamma^{I}e_{I}^{\mu}D_{\mu}\lambda +
D_{\mu}(e\gamma^{I}e_{I}^{\mu}\lambda) ]=0. \label{lambdabar2}
\end{eqnarray}
So if we impose the relation  $\varepsilon=\frac{\alpha}{2}$, the dynamical system we obtain is equivalent to the standard Palatini formulation of GR minimally coupled to
fermions. We therefore adopt that relation between the two parameters from here onwards.
In Ref.\cite{1stpaper}, the Hamiltonian analysis of the action (\ref{action}) without the matter part was carried out. In that case,
the Lagrangian equations of motion showed that torsion was zero on-shell although the action has
explicit torsion terms. In the next section we will carry out a complete Hamiltonian analysis with action (\ref{action}) where the torsion is expected to be non-zero.


\section{Hamiltonian analysis} \label{sec3}

We shall perform the Hamiltonian analysis of action (\ref{action}) similar to what was done in Ref.\cite{1stpaper} for the action without the matter term. Recall that in the Hamiltonian formulation of Hilbert-Palatini theory the basic variables are the $SO(1,3)$ spin connection $\omega_a^{~IJ}$
and its conjugate momentum. It is well known that this formulation contains second-class constraints. Since our action (\ref{action}) contains the other term which explicitly depends on torsion, we expect that there will be another pair of conjugate variables and the
second-class constraints will be somehow different from the Hilbert-Palatini case. It is also well known that in the absence of fermionic matter, torsion is zero. In the analysis of Ref.\cite{1stpaper}, this was obtained after we identified all the constraints. Owing to
the presence of the fermion term in the action, here torsion will not be zero. In this section we will show how the torsion and the spinorial degrees of freedom are related.

\subsection{3+1 Decomposition}

To seek a complete Hamiltonian analysis, we assume the spacetime be topologically $\Sigma\times \mathbb{R}$ with some compact spatial manifold $\Sigma$ without boundary so that the surface terms can be neglected. We first perform the $3+1$ decomposition of our fields
without breaking the internal $SO(1,3)$ symmetry and also without fixing any gauge. To identify our configuration and momentum variables for performing Hamiltonian analysis, we can rewrite the three pieces in the action as:
 \begin{eqnarray}
 S_{HP} &=& \int {\mbox d}^4
x \bigg[  e e^t_{[I}e^a_{J]} \left(\partial_{t} \omega_a^{~IJ}\right) + e e^t_{[I}e^a_{J]} \left(-\partial_{a} \omega_t^{~IJ} +  \omega_{[t}^{~IK} \omega_{a]}^{~KJ}\right)  + \frac{1}{2} e e^{a}_{[I} e^{b}_{J]} R_{a b}^{~~ IJ} \bigg] , \label{HP}\\ \alpha S_{T} &=&
\alpha\int {\mbox d}^4 x \bigg[\epsilon^{abc} D_{b}e_{c}^I \left(\partial_{t} e_a^I\right) + \epsilon^{abc} D_{b}e_{c}^I \left(-\partial_{a} e_t^I + \omega_{[t}^{~~IJ}e_{a]J} \right)\bigg] ,\label{torsion} \\ S_M &=& \int {\mbox d}^4 x ~ i e\bigg[ \left(
\overline{\lambda}(1 + i \frac{\alpha}{2}  \gamma_5) \gamma^t \partial_t \lambda - (\partial_t\bar{\lambda})\gamma^t (1 + i \frac{\alpha}{2} \gamma_5) \lambda \right) \nonumber\\ && \hspace{4em} + \frac{1}{2}\left( \overline{\lambda}(1 + i \frac{\alpha}{2} \gamma_5)
\gamma^t \omega_t^{IJ} \sigma_{IJ}\lambda +\overline{\lambda} \omega_t^{IJ} \sigma_{IJ} \gamma^t (1 + i \frac{\alpha}{2}  \gamma_5) \lambda \right)  \nonumber\\ && \hspace{4em}
 + \left( \overline{\lambda}(1 + i \frac{\alpha}{2}
\gamma_5) \gamma^a D_a \lambda - \overline{D_a \lambda}\gamma^a (1 + i \frac{\alpha}{2} \gamma_5) \lambda \right)\bigg].\label{matter}
\end{eqnarray}
We can read off the momenta with respect to $\omega_a^{~IJ}$, $e_a^I$, $\lambda$ and $\overline{\lambda}$ respectively
as
\begin{eqnarray}
\Pi^a_{IJ} :=  e e^t_{[I}e^a_{J]} ~~~~&,&~~~~ \Pi^a_I := \alpha \epsilon^{abc} D_{b}e_{c I} ,\label{momentumdef1} \\ \overline{\Pi} :=  i  e \overline{\lambda} (1 + i \frac{\alpha}{2} \gamma_5) e^t_I \gamma^I ~~~~&,&~~~~ \Pi := -i e  e^t_I \gamma^I
(1 + i \frac{\alpha}{2} \gamma_5) \lambda , \label{mattermomentum1}
 \end{eqnarray}
 where $\epsilon^{abc}$ denotes the 3-dimensional Levi-Civita tensor density, and we have used  the relation $\gamma^\mu = \gamma^I e^\mu_I$. For our analysis we shall use a standard
parametrization of the tetrad and the co-tetrad fields as in Ref.\cite{peldan}. This is the same parametrization used in the Hamiltonian analysis of the first two terms of our action in Ref.\cite{1stpaper}. Since the parametrization which we are using is standard, its
details and some related identities are given in Appendix \ref{app2}.

After some manipulation and neglecting the total derivatives, the pieces (\ref{HP}), (\ref{torsion}), and (\ref{matter}) of the action can be written in this parametrization respectively as
 \begin{eqnarray} S_{HP} &=& \int {\mbox d}^4 x \bigg[\Pi^{a}_{IJ} \partial_t
\omega_a^{~IJ} - \left( \frac{N^2}{2 e} \Pi^{[a}_{IK} \Pi^{b]}_{JL} \eta^{KL}  R_{a b}^{~~ IJ}  + \frac{1}{2}N^{[a}\Pi^{b]}_{IJ}  R_{a b}^{~~ IJ} -\omega_t^{~IJ}D_a \Pi^{a}_{IJ}  \right) \bigg] ,\label{HP2}\\ \alpha S_{T} &=& \int {\mbox d}^4 x \bigg[ \Pi^a_I \partial_t
V_a^I + \left( N N^I D_a \Pi^a_I +N^a V_a^I D_b \Pi^b_I + \frac{1}{2}\omega_t^{~IJ} \Pi^a_{[I} V_{J]a} \right) \bigg] ,\label{torsion2} \\ S_M  &=& \int {\mbox d}^4 x  \bigg[ \overline{\Pi} \partial_t \lambda  +  (\partial_t\bar{\lambda})\Pi   - \left( \frac{N}{\sqrt{q}}
\Pi^{a}_{IJ} \left( \overline{\Pi} \sigma^{IJ} D_a \lambda - \overline{D_a \lambda} \sigma^{IJ} \Pi \right) \right. \nonumber \\ && \hspace{8em} \left. + N^a \left( \overline{\Pi} D_a \lambda + \overline{D_a \lambda} \Pi \right) + \frac{1}{2} \omega_t^{~IJ} \left(
\overline{\lambda} \sigma_{IJ} \Pi - \overline{\Pi} \sigma_{IJ} \lambda \right) \right)  \bigg].\label{matter2}
\end{eqnarray}

\subsection{Primary and Secondary Constraints}

Let us now consider the constraints in the theory. At this stage we have the following constraints
\begin{enumerate}[(i)] \item Since there is no momentum corresponding to $\omega_t^{~IJ}$, we have to impose 6 primary constraints $\Pi^t_{IJ} \approx 0$. \item Also there
is no momentum corresponding to $e_t^I$. We have to impose 4 primary constraints $\Pi^t_{I} \approx 0$. From Eq.(\ref{newparameters1}) it is easy to see that this condition implies that there are no momenta corresponding to the lapse function $N$ and shift vector $N^a$.
Hence it will equivalently impose 4 primary constraints $\Pi_{N} \approx 0$ and $\Pi_{N^{a}} \approx 0$ . \item From Eq. (\ref{momentumdef1}), we can get two other sets of primary constraints
 \begin{eqnarray}
 C^a_I &:=& \Pi^a_I - \alpha \epsilon^{abc} D_{b}V_{c I}
\approx 0 , \label{newconstraint1}\\ \Phi^a_{IJ} &:=& \Pi^a_{IJ} -\frac{1}{2} \epsilon^{abc} \epsilon_{IJKL} V_b^K V_c^L \approx 0. \label{newconstraint2}
\end{eqnarray}
From Eq.(\ref{newconstraint1}) we get 12 constraints, while Eq.(\ref{newconstraint2}) gives 18
because of the antisymmetry in $IJ$. \item From the definition of the momenta corresponding to the fermions (Eq. (\ref{mattermomentum1})) we get 8 further constraints
\begin{eqnarray}
\Psi &:=& \Pi - i \sqrt{q}  N_K \gamma^K \left(1 + i \frac{\alpha}{2} \gamma_5\right)
\lambda \approx 0, \nonumber \\ \overline{\Psi} &:=& \overline{\Pi} +  i \sqrt{q} \overline{\lambda} \left(1 + i \frac{\alpha}{2} \gamma_5 \right) N_K \gamma^K  \approx 0. \label{newconstraint3}
\end{eqnarray}
\end{enumerate}

These are the primary constraints of our theory. By performing Legendre transformation, the Hamiltonian corresponding to the action (\ref{action}) can be expressed as
\begin{eqnarray}
H'&=&\int_{\Sigma}d^3x[\Pi^{a}_{IJ} \partial_t  \omega_a^{~IJ} + \Pi^a_I \partial_t
V_a^I + \overline{\Pi} \partial_t \lambda  + (\partial_t\bar{\lambda})\Pi -\mathcal{L}] \nonumber\\ &=&  \int_{\Sigma}d^3x(N H +N^a H_a + \omega_t^{~IJ} \mathcal{G}_{tIJ}),
\end{eqnarray}
where
\begin{eqnarray}
 H &=& \frac{1}{\sqrt{q}} \Pi^{a}_{IK}
\Pi^{b}_{JL} \eta^{KL} R_{a b}^{~~ IJ}  -  N^I D_a \Pi^a_I + \frac{1}{\sqrt{q}} \Pi^{a}_{IJ} \left( \overline{\Pi} \sigma^{IJ} D_a \lambda - \overline{D_a \lambda} \sigma^{IJ} \Pi \right) ,\label{hamiltonian1}\\ H_a &=&  \Pi^{b}_{IJ}  R_{a b}^{~~ IJ} - V_a^I D_b \Pi^b_I
+ \overline{\Pi} D_a \lambda + \overline{D_a \lambda} \Pi  , \label{diffeo1}\\
 \mathcal{G}_{tIJ} &=& - D_a \Pi^{a}_{IJ} - \frac{1}{2}\Pi^a_{[I} V_{J]a} +
 \frac{1}{2}\left( \overline{\lambda} \sigma_{IJ} \Pi - \overline{\Pi} \sigma_{IJ} \lambda\right) . \label{gauss1}
\end{eqnarray}
Subsequently we will drop the subscript $t$ from $\mathcal{G}_{tIJ}$ and denote it as $\mathcal{G}_{IJ}$. Including all of above primary constraints we can write the total Hamiltonian as
\begin{eqnarray}
 H_T := \int_{\Sigma}d^3x (N H + N^a H_a +
\omega_t^{~IJ} \mathcal{G}_{IJ} +\rho\Pi_{N}+\rho^{a}\Pi_{N^{a}}
 +\lambda_t^{IJ} \Pi^t_{IJ}+ \gamma_a^I C^a_I + \lambda_a^{IJ} \Phi^a_{IJ} + \overline{u}\Psi +  \overline{\Psi}u)
\label{htotal1},
\end{eqnarray}
where $\rho$,$\rho^{a}$,$\lambda_t^{IJ}$,$\gamma_a^I $,$\lambda_a^{IJ}$, $u$ and $\overline{u} $ are the Lagrangian multipliers.  At this point they are completely arbitrary. In order to preserve primary constraints $\Pi_{N} \approx 0$,
$\Pi_{N^{a}} \approx 0$ and $\Pi^t_{IJ} \approx 0$, one has to impose the following secondary constraints:
\begin{eqnarray*}
\dot{\Pi}_{N}&=&\{\Pi_{N}, H_T\}\approx0\Rightarrow H\approx0, \nonumber\\ \dot{\Pi}_{N^{a}}&=&\{\Pi_{N^{a}}, H_T\}\approx0 \Rightarrow
H_{a}\approx0, \nonumber\\ \dot{\Pi}^t_{IJ}&=&\{\Pi^t_{IJ} ,H_T\}\approx0 \Rightarrow \mathcal{G}_{IJ}\approx0, \nonumber
\end{eqnarray*}
which are called scalar,vector and Gaussian constraints respectively.

We now need to check whether the Hamiltonian system is consistent. To ensure the consistency of the Hamiltonian system, the constraints have to be preserved under evolution. Note that the primary constraints $\Pi_{N}$, $\Pi_{N^{a}}$ and $\Pi^t_{IJ}$ are preserved in
evolution respectively by the secondary constraints $H$, $H_{a}$ and $\mathcal{G}_{IJ}$. Note also that the Gaussian constraint $\mathcal{G}_{IJ}$ generates the $SO(1,3)$ transformations, and hence the Poisson bracket of any constraint with $\mathcal{G}_{IJ}$ is weakly
equal to zero. However, as shown in Ref.\cite{1stpaper} the constraint which actually generates the spatial diffeomorphisms for the gravitational variables is a combination given by
\begin{eqnarray}
\tilde{H_a}:= H_a +  \omega_a^{~IJ} \mathcal{G}_{IJ} +
\frac{1}{\alpha} \epsilon_{abc} C^b_I \Pi^c_I. \label{spatialdiffeo}
\end{eqnarray}
 This can be easily demonstrated as:
 \begin{eqnarray}
  \delta^{\tilde{H_a}} \omega_c^{~IJ}  &:=& \left\{ \omega_c^{~IJ},\tilde{H_a}(\nu^a)\right\} = \nu^a \partial_a \omega_c^{~IJ} +
\omega_a^{~IJ} \partial_c \nu^a = \mathcal{L}_{\nu^a}\omega_c^{~IJ} ,\nonumber \\ \delta^{\tilde{H_a}} \Pi^c_{IJ} &:=& \left\{ \Pi^c_{IJ}, \tilde{H_a}(\nu^a)\right\} = \nu^a \partial_a \Pi^c_{IJ} - \Pi^a_{IJ} \partial_a \nu^c +  \Pi^c_{IJ} \partial_a \nu^a   =
\mathcal{L}_{\nu^a}\Pi^c_{IJ} ,\nonumber \\ \delta^{\tilde{H_a}} V_c^I &:=&  \left\{  V_c^I,\tilde{H_a}(\nu^a)\right\} = \nu^a \partial_a  V_c^I +  V_a^I \partial_c \nu^a = \mathcal{L}_{\nu^a} V_c^I  ,\nonumber \\ \delta^{\tilde{H_a}}\Pi^c_I &:=&  \left\{ \Pi^c_I,
\tilde{H_a}(\nu^a)\right\} = \nu^a \partial_a \Pi^c_I  - \Pi^a_I \partial_a \nu^c + \Pi^c_I \partial_a \nu^a = \mathcal{L}_{\nu^a}\Pi^c_{I},
\end{eqnarray}
 where $\tilde{H_a}(\nu^a)\equiv \int_{\Sigma}d^3x \nu^a\tilde{H_a}$ denotes the smeared constraint. From now on, we
will keep this convention to denote the smeared version of a constraint with a smearing function, e.g., $\Psi(\overline{u})\equiv\int_{\Sigma}d^3x \overline{u} \Psi$. Also we will continue using the same notation $\omega_t^{~IJ}$ and $\gamma_a^I$ for the Lagrange
multipliers of $\mathcal{G}_{IJ}$ and $C^a_I$ respectively.

For the matter variables the constraint (\ref{spatialdiffeo}) acts as
\begin{eqnarray}
\delta^{\tilde{H_a}} \lambda = \left\{ \lambda, \tilde{H_a}(\nu^a)\right\} = \nu^a \partial_a \lambda ~&,&~ \delta^{\tilde{H_a}} \overline{\Pi} = \left\{ \overline{\Pi},
\tilde{H_a}(\nu^a)\right\} = \nu^a \partial_a \overline{\Pi} + \overline{\Pi} ~\partial_a \nu^a, \nonumber \\ \delta^{\tilde{H_a}}  \overline{\lambda} = \left\{ \overline{\lambda}, \tilde{H_a}(\nu^a)\right\} = \nu^a \partial_a \overline{\lambda} ~&,&~
\delta^{\tilde{H_a}} \Pi = \left\{ \Pi, \tilde{H_a}(\nu^a)\right\} = \nu^a \partial_a \Pi + \Pi ~\partial_a \nu^a.
\end{eqnarray}
Clearly this  combination $\tilde{H_a}$, acting on all the variables, generates Lie derivatives \cite{shyamreview} and can therefore be
identified as the diffeomorphism constraint. Using the property of Lie derivatives (or by explicit calculation) it can be shown that the Poisson bracket of any constraint with $\tilde{H_a}$ vanishes on the constraint surface. In fact we have
\begin{eqnarray*}
\{\tilde{H_b}(\mu^{b}),\tilde{H_a}(\nu^a)\}&=&\tilde{H_b}(-\mathcal{L}_{\nu^a}\mu^{b}),\\ \{H(M),\tilde{H_a}(\nu^a)\}&=&H(-\mathcal{L}_{\nu^a}M),\\ \{\Phi^b_{IJ}(\lambda_b^{IJ}),\tilde{H_a}(\nu^a)\}&=&\Phi^b_{IJ}(-\mathcal{L}_{\nu^a}\lambda_b^{IJ}),\\ \{
C^b_I(\gamma_b^I),\tilde{H_a}(\nu^a)\}&=&C^b_I(-\mathcal{L}_{\nu^a}\gamma_b^I),\\ \{\Psi(\overline{u}),\tilde{H_a}(\nu^a)\}&=&\Psi(-\mathcal{L}_{\nu^a}\overline{u}),\\ \{\overline{\Psi}(u),\tilde{H_a}(\nu^a)\}&=&\overline{\Psi}(-\mathcal{L}_{\nu^a}u).
\end{eqnarray*}
Note that the smeared scalar constraint reads $H(M)\equiv\int_{\Sigma}d^3x MH$ with $M$ as a smearing function. Now the $H_{a}$ term in the total Hamiltonian (\ref{htotal1}) can be replaced by $\tilde{H_a}$. Thus we can rewrite our total Hamiltonian as
\begin{eqnarray}
H_T := \int_{\Sigma}d^3x (N H + N^a \tilde{H_a} + \omega_t^{~IJ} \mathcal{G}_{IJ} + \gamma_a^I C^a_I + \lambda_a^{IJ} \Phi^a_{IJ} + \overline{u}\Psi +  \overline{\Psi}u + +\rho\Pi_{N}+\rho^{a}\Pi_{N^{a}}
 +\lambda_t^{IJ} \Pi^t_{IJ})
\label{htotal2}.
\end{eqnarray}

\subsection{Consistency Conditions}

The terms in the constraint algebra which are not weakly zero are respectively
\begin{eqnarray}
\left\{\Phi^a_{IJ}(\lambda_a^{IJ}), H(M) \right\} &=& \int_{\Sigma}d^3x\left(\frac{ M N_I \Pi^a_J}{\alpha} - \frac{\sqrt{q} }{2}M \overline{\lambda} \gamma_5 V^{a}_I \gamma_J
\lambda \right) \left(\alpha \lambda_a^{IJ} +\epsilon_{IJKL}\lambda_a^{KL}\right) ,\label{constalg1}\\
\left\{ C^a_I(\gamma_a^I),\Phi^b_{JK}(\lambda_b^{JK}) \right\} &=&\int_{\Sigma}d^3x \epsilon^{abc} \gamma_b^I V_c^J\left(\alpha \lambda_a^{IJ}
+\epsilon_{IJKL}\lambda_a^{KL}\right), \label{constalg2}\\
\left\{ C^a_I(\gamma_a^I), H(M) \right\} &=& -\int_{\Sigma}d^3x\frac{\alpha M}{\sqrt{q}} \epsilon^{abc} \gamma_b^I V_c^J \left( \overline{D_a \lambda} ~\sigma_{IJ} \Pi - \overline{\Pi} ~\sigma_{IJ} D_a \lambda
\right), \label{constalg3} \\
\left\{ C^a_I(\gamma_a^I), \overline{\Psi} (u) \right\} &=& -\int_{\Sigma}d^3x i  ~ \overline{\lambda} \left(1 + i \frac{\alpha}{2} \gamma_5 \right) \gamma^J \gamma_a^I~ \Pi^a_{IJ}u, \label{constalg4} \\
\left\{ C^a_I(\gamma_a^I),
\Psi(\overline{u}) \right\}  &=&\int_{\Sigma}d^3x i \overline{u} ~\gamma^J \left(1 + i \frac{\alpha}{2} \gamma_5\right) \lambda\gamma_a^I ~\Pi^a_{IJ}, \label{constalg5} \\
\left\{ \overline{\Psi} (u), \Psi(\overline{u}) \right\}  &=&\int_{\Sigma}d^3x 2 i\overline{u}
\sqrt{q} \gamma^I N_I u.  \label{constalg6}
\end{eqnarray}
For a consistent Hamiltonian system, the constraints should be preserved under evolution, i.e., for all the constraints $C_m$, we require $\dot{C_m} := \left\{ C_m, H_T \right\} \approx 0$. Our analysis will be
along the lines of Ref.\cite{1stpaper}. However, owing to presence of fermions, it will turn out that torsion is not zero. As a consequence, the calculations are much more complicated.

Let us first consider the consistency of constraint $\Phi^a_{IJ}$. From Eqs. (\ref{constalg1}) and (\ref{constalg2}) we need
\begin{eqnarray}
\dot{\Phi}^a_{IJ}(\sigma_a^{IJ}) &:=& \left\{\Phi^a_{IJ}(\sigma_a^{IJ}),H_T \right\}
 \nonumber\\
&\approx & \left\{\Phi^a_{IJ}(\sigma_a^{IJ}),H(N)\right\} + \left\{\Phi^a_{IJ}(\sigma_a^{IJ}) , C^b_I(\gamma_b^I)\right\} \approx 0  \label{phievol1}
\end{eqnarray}
where $\sigma_a^{IJ}$ is an arbitrary smearing function. Using Eqs. (\ref{constalg1}) and (\ref{constalg2}), and
after some calculation, Eq.(\ref{phievol1}) implies
\begin{eqnarray}
 && - \epsilon^{ade} \gamma_d^{[I} V^{J]}_e + \left( \frac{N N^{[I}\Pi^{a J]}}{\alpha} - \frac{\sqrt{q}}{2}N \overline{\lambda} \gamma_5 V^{a [I} \gamma^{J]} \lambda \right) \approx 0. \label{Phievo11}
\end{eqnarray}
Multiplying Eq.(\ref{Phievo11}) with $\epsilon_{abc}$, we have
\begin{eqnarray} \left(\gamma_b^I V_c^J -\gamma_c^I V_b^J - \gamma_b^J V_c^I +\gamma_c^J V_b^I  \right) - \epsilon_{abc}\frac{N}{\alpha}\left(N^I \Pi^{a J} - N^J \Pi^{a I}  - \frac{\alpha
\sqrt{q}}{2}\overline{\lambda} \gamma_5 \left[V^{a I} \gamma^J - V^{a J} \gamma^I \right] \lambda \right) \approx 0. \label{phievol2}
\end{eqnarray}
 Multiplying Eq.(\ref{phievol2}) with $V^b_J$ and using the properties (\ref{parameterconstraints2}) we get
\begin{eqnarray} 2 \gamma_c^I +V^b_J\gamma_b^J V_c^I - V^b_J \gamma_c^J V_b^I + \epsilon_{abc}\frac{N}{\alpha} N^I \Pi^{a J}V^b_J - \epsilon_{abc}\frac{N \sqrt{q}}{2} \overline{\lambda} \gamma_5 \left( V^{a I} V^b_J \gamma^J - V^{a J} V^b_J \gamma^I \right) \lambda
\approx 0. \label{phievol3}
\end{eqnarray}
By multiplying this equation with $N_I$, $V^c_I$ and $V_d^I$ respectively and using the relations (\ref{parameterconstraints1}) and (\ref{parameterconstraints2}), we obtain the following relations
\begin{eqnarray} \gamma_c^I
N_I &=& \frac{N}{2\alpha} \epsilon_{abc} \Pi^a_J V^b_J ,\label{phievolrel1}\\ \gamma_c^I V^c_I &=& 0 \label{phievolrel2} ,\\ \gamma_c^I V_{d I} &=& \epsilon_{dbc}\frac{N\sqrt{q}}{2} \overline{\lambda} \gamma_5 V^b_I \gamma^I \lambda, \label{phievolrel3}
\end{eqnarray}
where we have used Eq. (\ref{phievolrel2}) to obtain Eq. (\ref{phievolrel3}). Finally from Eqs. (\ref{phievolrel1}) and (\ref{phievolrel3}) we get a solution for the Lagrangian multiplier $\gamma_c^I$ as
\begin{eqnarray}
\gamma_c^I &=& \epsilon_{abc}\frac{N
\sqrt{q}}{2} \overline{\lambda} \gamma_5 V^a_I V^b_J \gamma^J \lambda - \frac{N}{2\alpha} \epsilon_{abc} N^I \Pi^a_J V^b_J. \label{gammasoln1}
\end{eqnarray}
Note that, all these equations differ from the corresponding equations in Ref.\cite{1stpaper} only by the
fermion-dependent terms. So, we have obtained 12 components of $\gamma_a^I$ from the 18 equations in Eq.(\ref{Phievo11}). Consequently there are 6 constraints remaining. By inserting the solutions (\ref{gammasoln1}) back into Eq.(\ref{Phievo11}) and after some
calculation, we get the following secondary constraint:
\begin{eqnarray}
\chi^{ab} := \Pi^a_{I} V^{b}_I + \Pi^b_{I} V^a_I - \alpha \sqrt{q} \overline{\lambda} \gamma_5 V^a_I V^b_I N_K \gamma^K \lambda \approx 0. \label{newconstraint4}
\end{eqnarray}
Since $\chi^{ab}$
is symmetric in $(a \leftrightarrow b)$, it contains just the 6 required constraints.

As seen above, the condition $\dot{\Phi}^a_{IJ} \approx 0$ fixed the Lagrange multipliers $\gamma_a^I$ of the constraint $C^a_I$ to the form given by Eq. (\ref{gammasoln1}). This can however be further simplified. For this and for subsequent calculations, we now derive
some useful identities using the constraint equations. All these identities hold weakly, i.e., they are true only when the constraints are used. From the definition of $\sigma_{IJ}$ and using the properties of gamma matrices (\ref{gammatprop1}), the Gaussian constraint
can also be written as
\begin{eqnarray}
\mathcal{G}_{IJ} = D_a \Pi^{a}_{IJ} +  \frac{1}{2}\Pi^a_{[I} V_{J]a} + \frac{\sqrt{q}}{4} \left( \alpha \overline{\lambda} \gamma_5 N_{[I} \gamma_{J]}\lambda + \epsilon_{IJKL}  \overline{\lambda} \gamma_5 N^{[K} \gamma^{L]}
\lambda  \right) \approx 0. \label{gauss2}
\end{eqnarray}
From the constraints (\ref{newconstraint1}) and (\ref{newconstraint2}) we can easily obtain the relation:
\begin{eqnarray}
D_{IJ} ~:= ~ D_a \Pi^a_{IJ} -  \frac{1}{\alpha}\epsilon_{IJKL} \Pi^{a K} V_a^L ~
\approx 0 ~ .\label{defDIJ}
\end{eqnarray}
 Using this and the Gaussian constraint (\ref{gauss2}) we get, after some algebra,
 \begin{eqnarray}
 \Pi^a_{[I}V_{J] a} + \frac{\alpha\sqrt{q}}{2} \overline{\lambda} \gamma_5 N_{[I} \gamma_{J]}\lambda ~ \approx ~ 0.
\label{identityA}
\end{eqnarray}
Multiplying this equation with $N_J$ and then with $V^b_I$, and using the properties (\ref{parameterconstraints1}) and (\ref{parameterconstraints2}), we get
\begin{eqnarray}
\Pi^b_J N^J ~ \approx ~ \frac{\alpha\sqrt{q}}{2}
\overline{\lambda} \gamma_5 \gamma_J V^{b J} \lambda. \label{identityA1}
\end{eqnarray}
By multiplying relation (\ref{identityA}) with $V^b_I$ and then with $V^c_J$, and again using the properties (\ref{parameterconstraints1}) and (\ref{parameterconstraints2}), we get
\begin{eqnarray}
\Pi^c_I V^b_I - \Pi^b_I V^c_I ~ \approx ~ 0. \label{identityA2}
\end{eqnarray}
Plugging Eq. (\ref{identityA2}) in the constraint (\ref{newconstraint4}) we get the relation
\begin{eqnarray}
\Pi^a_{I} V^{b}_I ~ \approx ~  \frac{\alpha \sqrt{q}}{2}
\overline{\lambda} \gamma_5 V^a_I V^b_I N_K \gamma^K \lambda. \label{identityA3}
\end{eqnarray}
These identities can be used to greatly simplify the subsequent calculations.

Note that because of the identity (\ref{identityA2}), the second term on the RHS of Eq. (\ref{gammasoln1}) drops out and the Lagrangian multiplier of $C^c_I$ in $H_T$ becomes
\begin{eqnarray}
\gamma_c^I &=& \epsilon_{abc}\frac{N \sqrt{q}}{2} \overline{\lambda} \gamma_5
V^a_I V^b_J \gamma^J \lambda . \label{gammasoln2}
\end{eqnarray}
This leads to further simplification of our problem. Moreover, let us consider the identity (\ref{identityA3}) again. Multiplying it by $V_{b I}$ and using Eq.(\ref{identityA1}), properties
(\ref{parameterconstraints2}) and (\ref{identity2}) , we get
\begin{eqnarray}
\Pi^a_{I} ~ \approx ~ \frac{\alpha \sqrt{q}}{2} \overline{\lambda} \gamma_5 \left(V^a_{[I} N_{J]} \right) \gamma^J \lambda ~ = ~\frac{\alpha}{2} \overline{\lambda} \gamma_5 \Pi^a_{IJ}
\gamma^J \lambda. \label{torspinreln}
\end{eqnarray}
This equation relates the torsion degrees of freedom encoded in $\Pi^a_{I}$ with the spin degrees of freedom $\lambda$ and $\overline{\lambda}$. Note that we have used only constraint equations and not equations of
motion in deriving Eq.(\ref{torspinreln}). This is a weak relation since it has been derived by using the constraints $\mathcal{G}_{IJ},C^a_I,\Phi^a_{IJ},\chi^{ab}$. When there is no matter, this equation would indicate that torsion is zero \cite{1stpaper}. Note also
that relation (\ref{torspinreln}) is as same as the one obtained in Ref.\cite{shyam}.

Now let us consider the constraints $\Psi$ and $\overline{\Psi}$. For the consistency conditions for constraints $\Psi$ and $\overline{\Psi}$, we need
\begin{eqnarray}
 \dot{\Psi}(\overline{v}):=\{\Psi(\overline{v}), H_T\}= \int_{\Sigma}d^3x[-i \overline{v} ~\gamma^J
\left(1 + i \frac{\alpha}{2} \gamma_5\right) \lambda\gamma_a^I ~\Pi^a_{IJ}-2 i\overline{v} \sqrt{q} \gamma^I N_I u]\approx 0,\  \forall\overline{v}. \label{dotPhi}
\end{eqnarray}
Note that since $\chi^{ab}$ is a secondary constraint, we do not add
it in $H_{T}$. As proved beforehand, the condition that
$\Phi^a_{IJ}$ be preserved under evolution has fixed $\gamma_a^I$ to the specific form given by Eq. (\ref{gammasoln2}). Now recall from Eq. (\ref{constalg5}), for an arbitrary smearing function $\xi_a^I$ we have
\begin{eqnarray}
\left\{ C^a_I(\xi_a^I),
\Psi(\overline{v}) \right\}  &=&\int_{\Sigma}d^3x i \overline{v} ~\gamma^J \left(1 + i \frac{\alpha}{2} \gamma_5\right) \lambda\xi_a^I ~\Pi^a_{IJ}, \nonumber
\end{eqnarray}
When $\xi_a^I  =  \gamma_a^I$, which is of the form given in Eq. (\ref{gammasoln2}), using Eqs.
(\ref{identity2}) and (\ref{parameterconstraints1}) we get
\begin{eqnarray} \gamma_c^I~ \Pi^c_{IJ} ~ = ~\left( \epsilon_{abc}\frac{N q}{2} \overline{\lambda} \gamma_5 \gamma^K \lambda \right) V^a_I V^b_K V^c_{[I} N_{J]} = 0. \nonumber
\end{eqnarray}
Therefore, once
the Lagrange multiplier $\gamma_a^I$ is fixed to the value required for a consistent Hamiltonian system, Eq.(\ref{dotPhi}) becomes:
\begin{eqnarray}
 \dot{\Psi}(\overline{v}):=\{\Psi(\overline{v}), H_T\}=-\int_{\Sigma}d^3x 2 i\overline{v} \sqrt{q} \gamma^I N_I u\approx
0,\  \forall \overline{v}.\label{dotPhi2}
\end{eqnarray}
By using Eqs. (\ref{newparameters1}) and (\ref{parameterconstraints1}), we can obtain $\gamma_{I}N^{I}=\gamma^{\mu}e_{\mu I}N^{I}=-\gamma^{t}N$. Since both $\gamma^{t}$ and $N$ are nonzero, one has $\gamma_{I}N^{I}\neq0$.
It is obvious that the only solution for Eq.(\ref{dotPhi2}) is $u=0$. Similarly, we need
\begin{eqnarray}
\dot{\overline{\Psi}}(v)&:=&\{\overline{\Psi}(v), H_T\}=\int_{\Sigma}d^3x[i  ~ \overline{\lambda} \left(1 + i \frac{\alpha}{2} \gamma_5 \right) \gamma^J \gamma_a^I~
\Pi^a_{IJ}v+2 i\overline{u} \sqrt{q} \gamma^I N_I v]\nonumber\\ &=&\int_{\Sigma}d^3x2 i\overline{u}\sqrt{q} \gamma^I N_I v\approx 0,\ \forall v.
\end{eqnarray}
Its only solution is $\overline{u}=0$.

We now turn to the additional secondary constraint $\chi^{ab}$ (see Eq.(\ref{newconstraint4})). We now have to check its contribution to the constraint algebra. Obviously, $\chi^{ab}$ commutes with primary constraints $\Pi_{N}$, $\Pi_{N^{a}}$ and $\Pi^t_{IJ}$. Moreover
one has
\begin{eqnarray*}
\{\chi^{ab}(\sigma_{ab}),\mathcal{G}_{IJ}(\Lambda^{IJ})\}&\approx&0,\\ \{\chi^{bc}(\sigma_{bc}),\tilde{H_a}(\nu^a)\}&=&\chi^{bc}(-\mathcal{L}_{\nu^a}\sigma_{bc}).
\end{eqnarray*}
The additional non-zero terms in the constraint algebra are
\begin{eqnarray}
\left\{\chi^{ab}(\sigma_{ab}), \Psi(\overline{u})  \right\} &=& \int_{\Sigma}d^3x2 i \overline{u} \sigma_{ab}\sqrt{q} V^a_I V^{b I} N_J \gamma^J \lambda, \label{constalg7}\\ \left\{\chi^{ab}(\sigma_{ab}), \overline{\Psi}(u) \right\} &=&
-\int_{\Sigma}d^3x2 i \sigma_{ab}\sqrt{q} V^a_I V^{b I} \overline{\lambda}\gamma^J N_J u, \label{constalg8}\\ \left\{\chi^{ab}(\sigma_{ab}), C^c_I(\gamma_c^I)\right\} &=&\int_{\Sigma}d^3x [\frac{\alpha \sigma_{ac}}{2 \sqrt{q}} \epsilon^{cdb} \gamma_d^{[I} V^{J]}_b (
\Pi^a_{[I} N_{J]} - \alpha\Pi^a_{IJ}  \overline{\lambda} \gamma_5 N_K \gamma^K \lambda)  \label{constalg9} \\
 && + \frac{\sigma_{ab}}{\sqrt{q}}\gamma_c^I N_I \left( 2 \Pi^a_K
V^c_J \Pi^b_{KJ} - \frac{\alpha}{2} \Pi^a_{JL} V^c_K  \Pi^b_{JL} \overline{\lambda} \gamma_5 \gamma^K \lambda  \right) - \frac{2 \alpha \sigma_{cb}}{\sqrt{q}}\epsilon^{cad} N_J \Pi^b_{IJ} D_a \gamma_d^I ], \nonumber\\ \left\{\Phi^c_{IJ}(\lambda_c^{IJ}),
\chi^{ab}(\sigma_{ab}) \right\} &=&\int_{\Sigma}d^3x
 2 \sigma_{ab} \lambda_c^{IJ} \epsilon^{acd} \epsilon_{IJKL} V^b_K V_d^L, \label{constalg10} \\
\left\{\chi^{ab}(\sigma_{ab}),H(M) \right\} &=&\int_{\Sigma}d^3x \{\frac{\sigma_{ac}}{\sqrt{q}}\left[ \Pi^a_{[I} N_{J]} - \alpha \Pi^a_{IJ}  \overline{\lambda} \gamma_5 N_M \gamma^M \lambda \right] \left[ D_b \left( \frac{M}{\sqrt{q}} \Pi^b_{[J|L|}  \Pi^c_{I]L}  \right)
- \frac{M}{2} N_{[I} \Pi^c_{J]} \right. \label{constalg11} \\ &+&\left.\frac{M \Pi^c_{KL} }{2 \sqrt{q}} \left( \overline{\lambda} \sigma^{IJ} \sigma^{KL} \Pi + \overline{\Pi} \sigma^{KL} \sigma^{IJ} \lambda \right) \right] - \frac{2 M}{q} \sigma_{cb} N_J \Pi^b_{IJ}
\Pi^c_{IK} D_a \Pi^a_K \nonumber \\ &+& \frac{M \alpha \sigma_{ab}}{2 q}\Pi^a_{IJ}\Pi^b_{IJ} \Pi^c_{ML} N_K \left(\overline{\lambda} \gamma_5 \gamma^K \sigma^{ML} D_c \lambda - \overline{D_c \lambda} \sigma^{ML}\gamma_5 \gamma^K \lambda \right) \nonumber \\ &+&
\frac{\sigma_{ab}}{2 \sqrt{q}} \left( 4 \Pi^a_K N_I V^c_J \Pi^b_{KJ} - \alpha \Pi^a_{JL} \Pi^b_{JL}  V^c_K  N_I \overline{\lambda} \gamma_5 \gamma^K \lambda \right) D_c\left(M N^I \right)\}. \nonumber
\end{eqnarray}
The consistency conditions of constraints $C^a_I$ and
$\chi^{ab}$ read respectively
 \begin{eqnarray} \dot{C}^a_I(\eta_a^I) &=& \left\{C^a_I (\eta_a^I), H_T\right\} = \left\{C^a_I (\eta_a^I), \left( \Phi^a_{IJ}( \lambda_a^{IJ}) + H(N)\right) \right\} \approx 0, \label{Cdot} \\
\dot{\chi}^{ab}(\sigma_{ab})  &=& \left\{\chi^{ab}(\sigma_{ab}),H_T\right\} = \left\{\chi^{ab}(\sigma_{ab}), \left( \Phi^a_{IJ}( \lambda_a^{IJ}) + C^a_I (\gamma_a^I) + H(N) \right)  \right\} \approx 0, \label{chidot}
\end{eqnarray}
where $\eta_a^I$ and $\sigma_{ab}$ are arbitrary smearing functions, $H_{T}$ is still given by Eq.(\ref{htotal2}). It turns out that we can indeed solve the 18 independent equations (\ref{Cdot}) and (\ref{chidot}) to fix the 18 independent components of the Lagrangian
multiplier $\lambda_a^{IJ}$. This calculation is slightly lengthy and complicated and has been given in Appendix (\ref{app3}).

We are finally left with the scalar constraint. We now need to prove $\dot{H}(M)\approx0$. The time evolution of scalar constraint reads
\begin{eqnarray}
\dot{H}(M)&:=&\{H(M),H_T \}=\{H(M),\Phi^a_{IJ}( \lambda_a^{IJ}) + C^a_I (\gamma_a^I)\} \nonumber\\
&=&-\int_{\Sigma}d^3x\left(\frac{ M N_I \Pi^a_J}{\alpha} - \frac{\sqrt{q} }{2}M \overline{\lambda} \gamma_5 V^{a}_I \gamma_J \lambda \right) \left(\alpha \lambda_a^{IJ} +\epsilon_{IJKL}\lambda_a^{KL}\right) \nonumber\\ && +\int_{\Sigma}d^3x\frac{\alpha M}{\sqrt{q}}
\epsilon^{abc} \gamma_b^I V_c^J \left( \overline{D_a \lambda} ~\sigma_{IJ} \Pi - \overline{\Pi} ~\sigma_{IJ} D_a \lambda \right), \label{Hevo}
\end{eqnarray}
 where $\gamma_a^I$ and $\lambda_a^{IJ}$ are given by Eq.(\ref{gammasoln2}) and Eq.(\ref{XaIJ2}) respectively. By
using Eq.(\ref{XaIJ1}), we have
\begin{eqnarray}
\dot{H}(M)&=&-\int_{\Sigma}d^3x\left(\frac{M N_I \Pi^a_J}{\alpha} - \frac{\sqrt{q} }{2}M\overline{\lambda} \gamma_5 V^{a}_I \gamma_J \lambda \right) [\frac{\alpha N}{\sqrt{q}} \left( \overline{D_a \lambda} ~\sigma_{IJ} \Pi
-\overline{\Pi} ~\sigma_{IJ} D_a \lambda \right)] \nonumber\\ && +\int_{\Sigma}d^3x\frac{\alpha M}{\sqrt{q}} \epsilon^{abc} \gamma_b^I V_c^J \left( \overline{D_a \lambda} ~\sigma_{IJ} \Pi - \overline{\Pi} ~\sigma_{IJ} D_a \lambda \right) \nonumber\\ &&
-\int_{\Sigma}d^3x\left(\frac{M N_I \Pi^a_J}{\alpha} - \frac{\sqrt{q} }{2} M \overline{\lambda} \gamma_5 V^{a}_I \gamma_J \lambda \right)X_a^{IJ}. \label{dotH1}
\end{eqnarray}
 Using the solution (\ref{gammasoln2}) of $\gamma_a^I$  and also using Eq.(\ref{torspinreln}),
it can be shown that the first two terms in Eq.(\ref{dotH1}) cancel each other. For the last term of above equation, by using Eq.(\ref{XaIJ3}) and properties (\ref{parameterconstraints1}) and (\ref{parameterconstraints2}) we find it is exactly equal to zero. Therefore we
get $\dot{H}(M)\approx0$. We have now exhausted all the consistency conditions. We have also proved that the constraints are preserved under evolution, i.e., for all the constraints $C_m$, we have shown $\dot{C_m} := \left\{ C_m, H_T \right\} \approx 0$. We have
therefore obtained a consistent Hamiltonian system.

Now all the constraints have been identified, we can classify them into first-class constraints and second-class ones. It is obvious that $\tilde{H}_{a}$ and $\mathcal{G}_{IJ}$ are first class. Since none of constraints contain $N$, $N^{a}$ or $\omega_t^{~IJ}$, primary
constraints $\Pi_{N} \approx 0$, $\Pi_{N^{a}} \approx 0$ and $\Pi^t_{IJ} \approx 0$ are first class. In this sense, $N$, $N^{a}$ and $\omega_t^{~IJ}$ are arbitrary Lagrangian multipliers. We may eliminate configuration $N$, $N^{a}$ and $\omega_t^{~IJ} $ as well as their
conjugate momenta $\Pi_{N}$, $\Pi_{N^{a}}$ and $\Pi^t_{IJ}$ from dynamical variables\cite{canbin,thiemannbook}. The term $\rho\Pi_{N}+\rho^{a}\Pi_{N^{a}}
 +\lambda_t^{IJ} \Pi^t_{IJ}$ in the total Hamiltonian (\ref{htotal2}) can be eliminated. Thus we get
\begin{eqnarray}
H_T := \int_{\Sigma}d^3x (N H + N^a \tilde{H_a} + \Lambda^{IJ} \mathcal{G}_{IJ} + \gamma_a^I C^a_I + \lambda_a^{IJ} \Phi^a_{IJ}) \label{htotal3},
\end{eqnarray}
 where $\omega_t^{~IJ}$ is replaced by $\Lambda^{IJ}$. In light of the argument given
above, $\Lambda^{IJ}$, $N$ and $N^a$ are just Lagrangian multipliers. At this stage, $\gamma_a^I$ and $\lambda_a^{IJ}$ in the above total Hamiltonian (\ref{htotal3}) are given by Eq.(\ref{gammasoln2}) and Eq.(\ref{XaIJ2}) respectively. Also we have removed the
second-class constraints $\Psi$ and $\overline{\Psi}$ from $H_T$ as we have proved that the Lagrange multipliers for these two constraints, $\overline{u}$ and $u$ respectively, are zero. Recall that all the non-zero terms of the constraint algebra are given in Eqs.
(\ref{constalg1}-\ref{constalg6}) and (\ref{constalg7}-\ref{constalg11}). It can be easily seen that $\Phi^a_{IJ}$, $C^a_I$, $\chi^{ab}$, are second class. Although the Poisson brackets of some constraints with the scalar constraint $H$ are still not weakly equal to
zero, it can be shown that we can construct a new first-class Hamiltonian constraint by the combination:
\begin{eqnarray}
 \tilde{H} = H + \frac{\gamma_a^I}{N} C^a_I + \frac{\lambda_a^{IJ}}{N} \Phi^a_{IJ}. \label{Htilde}
 \end{eqnarray}

Since we have already identified all the constraints, we can now count the degrees of freedom. The gravitational degrees of freedom are incorporated in the pair $\left( \Pi^a_{~IJ}, \omega_a^{~IJ} \right)$, which have 36 degrees of freedom and in the pair
$\left(\Pi^a_I,V_a^I  \right)$, which have 24. The matter degrees of freedom are in the two pairs $\left( \overline{\Pi}, \lambda\right)$ and $\left( \Pi, \overline{\lambda}\right)$ each with 8 degrees of freedom. The total number of degrees of freedom without
considering the constraints is therefore 76. Clearly $\mathcal{G}_{IJ}$, $\tilde{H_a}$ and $\tilde{H}$ contribute $(6+3+1)=10$ first-class constraints removing $20$ degrees of freedom. The constraints $C^a_I$, $\Phi^a_{IJ}$, $\Psi$ and $\overline{\Psi}$ are primary
second class removing $(12+18+4+4)=38$ degrees of freedom. Finally the secondary constraint $\chi^{ab}$ turns out to be second class and thus removes $6$ degrees of freedom. Thus the number of independent degrees of freedom in our system is $12$. In these $12$ degrees of
freedom, $4$ represent gravity and $8$ denote Dirac fermions.


\section{Solving the Second-Class Constraints}\label{sec4}

Second-class constraints are problematic because the flows generated by them do not lie on the constraint surface. Having obtained a consistent Hamiltonian system in the previous section, we now proceed to solve all the second-class constraints and eliminate spurious
degrees of freedom. We will do this  after performing a partial gauge fixing. Since we have already proved the consistency of the Hamiltonian system, we can be sure that making a gauge choice now will not lead to any inconsistency.

Our goal is to reduce the internal $SO(1,3)$ gauge symmetry to $SU(2)$. So we break the $SO(1,3)$ symmetry by fixing the internal timelike vector $N^I = \left( 1,0,0,0 \right)$, i.e.,we fix a specific timelike direction in the internal space. This is a standard gauge choice and
is known as \textit{time gauge}. From Eqs.  (\ref{newparameters1}), (\ref{parameterconstraints1}) and (\ref{parameterconstraints2}) it is easy to see that
\begin{eqnarray}
 N^I = \left( 1,0,0,0 \right) ~ ~ \Leftrightarrow ~ ~ V_a^0 ~ = ~ 0 ~ = ~ V^a_0. \label{timegauge}
\end{eqnarray}
For consistency, this gauge fixing condition has to be preserved, i.e.,
\begin{eqnarray}
\dot{V}_a^0 =  \left\{ V_a^0, H_T \right\} \approx 0. \nonumber
\end{eqnarray}
 Hence in time gauge we get
 \begin{eqnarray}
 \Lambda^{0i}~=~ V^{ai}\partial_a N.
\label{timegaugepreserve}
\end{eqnarray}
The Lagrangian multiplier $\Lambda^{0i}$ of $\mathcal{G}_{0i}$ gets fixed. This is expected because, by fixing $N^I$, we have broken the $SO(1,3)$ gauge invariance. The preservation of this gauge fixing condition implies that
the boost part of the Gaussian constraint does not generate gauge transformations.

We first solve the constraint (\ref{newconstraint2}), which can be written as
\begin{eqnarray}
\Pi^a_{IJ} = \frac{1}{2} \epsilon^{abc} \epsilon_{IJKL} V_b^K V_c^L. \nonumber
\end{eqnarray}
Thus in time gauge we have
\begin{eqnarray}
\Pi^a_{ij} = 0 \hspace{1em};
\hspace{1em} ~\Pi^a_{0i} ~= ~ \frac{1}{2}\epsilon^{abc} \epsilon_{ijk} V_b^j V_c^k :=  E^a_i. \label{phisoln}
\end{eqnarray}
 So, after solving this constraint only the $\Pi^a_{0i}$ part of $\Pi^a_{IJ}$ remains a basic dynamical variable. Consequently, only the
$\omega_a^{~0i}$ part of the $SO(1,3)$ connection remains basic dynamical variable. For convenience we define $K_a^i := 2 \omega_a^{~0i}$ which will be conjugate to $E^a_i$. The $\omega_a^{~ij}$ is the remaining part of the connection which will get solved in terms of
other variables while solving the remaining constraints. Since our gauge group is now reduced to $SU(2)$, we will expand  $SO(1,3)$ connection components $\omega_a^{~ij}$ in the adjoint basis of $SU(2)$ as $\omega_a^{~ij} := - \epsilon^{ijk} \Gamma_{a k}$. The quantity
$\Gamma_{a k}$ is the $SU(2)$ spin connection. Note that we had started with a $SO(1,3)$ spin connection $\omega_a^{~IJ}$ which was not torsion-free. Therefore the variables $K_a^i$ and $\Gamma_a^i$ that we define above will contain information about torsion implicitly.
Also, from Eq.(\ref{parameterconstraints2}) it is clear that, in time gauge, $V^a_i$ is the inverse of $V_a^i$. Using the properties of inverses and determinants of matrices, it is easy to see from Eq.(\ref{phisoln}) that $E^a_i = \sqrt{q} V^a_i$ is the densitized triad.
We can also determine its inverse $E_a^i = \frac{1}{\sqrt{q}}V_a^i$. \footnote{In this work, we are not interested in the behaviour under parity transformations. Therefore we omit the $\mbox{sgn}( \mbox{det} E^a_i)$ terms from our expressions.} Next we consider the
constraint (\ref{newconstraint4}). It can be easily seen that once we substitute (\ref{torspinreln}) into the expression of $\chi^{ab}$, it is identically satisfied. Using the above  solution (\ref{phisoln}), in time gauge the equation (\ref{torspinreln}) simplifies to
\begin{eqnarray}
\Pi^a_0 = \frac{\alpha}{2} E^a_i \overline{\lambda} \gamma_5 \gamma^i \lambda \hspace{1em}; \hspace{1em} \Pi^a_i = - \frac{\alpha}{2} E^a_i\overline{\lambda} \gamma_5 \gamma^0 \lambda .\label{chisoln}
\end{eqnarray}
The torsion degrees of freedom are
solved in terms of the densitized triad $E^a_i$ and the fermionic fields $\lambda$ and $\overline{\lambda}$.

Using above results we can now solve the second-class constraint (\ref{newconstraint1}) as
\begin{eqnarray}
C^a_0 = 0 ~ ~\Rightarrow ~ && \frac{E^a_i }{2} \left( \overline{\lambda} \gamma_5 \gamma^i \lambda - \frac{1}{\sqrt{q}} \epsilon_{ijk} K^j_b E^{b k} \right) ~ =
~ 0, \label{ca0sol} \\ C^a_i = 0 ~ ~\Rightarrow ~ &&  \frac{E^a_i}{2}\overline{\lambda} \gamma_5 \gamma^0 \lambda + \frac{1}{\sqrt{q}}\left( \epsilon^{jkl} E^a_j E^b_k E^c_l \partial_b E_c^i + \frac{1}{2} \epsilon^{ijk} E^a_j E^b_k  E_c^l \partial_b E^c_l \right) +
\frac{1}{\sqrt{q}} \bigg( \Gamma_b^k E^b_k E^a_i -  \Gamma_b^k E^b_i E^a_k \bigg)  = ~ 0. \nonumber \\ \label{caisol}
\end{eqnarray}
Equation (\ref{caisol}) can be used to solve the spin connection $\Gamma_a^i$ in terms of the other variables. After some algebra we
obtain
\begin{eqnarray}
\Gamma_a^i &=& \frac{1}{2}\epsilon^{ijk} E_a^j E^b_k E^c_l \partial_b E_c^l + \frac{1}{2}\epsilon^{ijk} E_a^l E^b_j E^c_k \partial_b E_c^l + \frac{1}{2}\epsilon^{ijk} E^b_k \partial_a E_b^j - \frac{1}{2}\epsilon^{ijk} E^b_k \partial_b E_a^j -
\frac{\sqrt{q}}{4} E_a^i\overline{\lambda} \gamma_5 \gamma^0 \lambda  \label{gammasol1} \\ &:=& ~ \widehat{\Gamma}_a^i - \frac{\sqrt{q}}{4} E_a^i \overline{\lambda} \gamma_5 \gamma^0 \lambda,  \label{gammasol2}
\end{eqnarray}
where we have denoted by
$\widehat{\Gamma}_a^i$ the first four terms in the RHS of Eq.(\ref{gammasol1}), which do not depend on the fermions. It turns out that $\widehat{\Gamma}_a^i$ is exactly the $SU(2)$ spin connection which we would have obtained, had there been no fermionic matter
\cite{thiemannbook}. So, when there is no matter we go back to the standard GR formulation. Also note that the spin connection $\Gamma_a^i$ is independent of the arbitrary coupling parameter $\alpha$.
So far, we have \textit{reduced} our original phase space by consistently imposing time gauge and then solving some second-class constraints. As a result, some basic variables in the original phase space have been eliminated in terms of the others. To obtain the basic
variables in this phase space we need to find the symplectic structure after all these reductions.

Recall that, we started with the symplectic structure given by:
\begin{eqnarray}
\int\left[\Pi^{a}_{IJ} \partial_t  \omega_a^{~IJ} + \Pi^a_I \partial_t V_a^I + \overline{\Pi} \partial_t \lambda  + (\partial_t \overline{\lambda}) \Pi  \right].  \label{symplectic1}
\end{eqnarray}
 Using Eq. (\ref{phisoln}) and our definition $K_a^i := 2 \omega_a^{~0i}$, the first term in expression (\ref{symplectic1}) becomes $E^a_i\partial_t K_a^i$.
For the second term, recall that  $\partial_t V_a^0 = 0$ in time gauge. Then
\begin{eqnarray}
\Pi^a_I \partial_t V_a^I&=&\Pi^a_i \partial_t V_a^i~. \nonumber
\end{eqnarray}
Using the constraint equation (\ref{newconstraint1}) we can calculate the second term:
\begin{eqnarray}
\int \Pi^a_i \partial_t V_a^i &=&\int\alpha \epsilon^{bca} \left(\partial_{b}V_{c i}\cdot\partial_t V_a^i+\omega_b^{~ij}V_{cj}\cdot\partial_t V_a^i \right) \nonumber\\
 &=&\int E^{a}_{i}\partial_t(-\alpha\Gamma^{i}_{a}) \nonumber\\
&=&\int [E^{a}_{i}\partial_t(-\alpha\widehat{\Gamma}_a^i)-\frac{\alpha}{4}\sqrt{q}\lambda^\dagger \gamma_5 \lambda E_a^i \partial_t E^{a}_{i}]  ,
\end{eqnarray}
where we also used the solution (\ref{phisoln}) and neglected total derivative terms.
For the last two terms, recall that $\overline{\lambda} := \lambda^\dagger \gamma^0$. Also $\Pi$ and $\overline{\Pi}$ can be read off from the constraints (\ref{newconstraint3}). Then in time gauge, using the properties of the $\gamma$ matrices given in appendix
(\ref{app1}) we get
\begin{eqnarray}
\overline{\Pi} \partial_t \lambda  + (\partial_t \overline{\lambda}) \Pi = -i \sqrt{q}  \lambda^\dagger \left( \partial_t \lambda \right) +  i \sqrt{q} \left( \partial_t \lambda^\dagger \right) \lambda + \frac{\alpha}{2}
\lambda^\dagger \gamma_5 \lambda ~ \partial_t \left(  \sqrt{q} \right) - \partial_t \left( \frac{\alpha}{2}  \sqrt{q}  \lambda^\dagger \gamma_5 \lambda \right). \nonumber
\end{eqnarray}
Since $\partial_t \sqrt{q} = \frac{1}{2}\sqrt{q}E_a^i \partial_t E^{a}_{i}$, we
have
\begin{eqnarray}
\int\left[\overline{\Pi} \partial_t \lambda  + (\partial_t \overline{\lambda}) \Pi \right] = \int \left[ -i \sqrt{q}  \lambda^\dagger \left( \partial_t \lambda \right) +  i \sqrt{q} \left( \partial_t \lambda^\dagger \right) \lambda
+\frac{\alpha}{4}\sqrt{q}\lambda^\dagger \gamma_5 \lambda E_a^i \partial_t E^{a}_{i} \right], \label{sym3rdterm}
\end{eqnarray}
where we have again neglected the total time derivative term.
Putting everything together, expression (\ref{symplectic1}) becomes
\begin{eqnarray}
 && \int \left[ E^a_i\partial_t (K_a^i-\alpha\widehat{\Gamma}_a^i) -i \sqrt{q}  \lambda^\dagger \left( \partial_t \lambda \right) + i \sqrt{q} \left( \partial_t \lambda^\dagger \right)
\lambda \right] \nonumber\\ && \equiv \int \left[E^a_i\partial_t (K_a^i-\alpha\widehat{\Gamma}_a^i) + \left( \partial_t {\zeta}^\dagger \right)  \Pi_{\zeta^\dagger} + \Pi_\zeta \left( \partial_t \zeta \right)\right], \label{symplectic2}
\end{eqnarray}
 where, following
Refs.\cite{thiemannfermion,martinfermion}, we have defined half-densities of the fermionic variables: $\zeta := \sqrt[4]{q} \lambda$, $\zeta^\dagger  := \sqrt[4]{q} \lambda^\dagger$ and identified $\Pi_\zeta = - i \zeta^\dagger$, $\Pi_{\zeta^\dagger} = i \zeta$.

The second-class constraints $\Psi$ and $\overline{\Psi}$ now become
\begin{eqnarray}
 \psi ~ := ~ \Pi_\zeta + i \zeta^\dagger ~ \approx 0 \hspace{1em} &&\hspace{1em}; \hspace{2em} \tilde{\psi} ~:= ~ \Pi_{\zeta^\dagger} - i \zeta ~ \approx 0. \end{eqnarray}
 The two
constraints $\psi$ and $\tilde{\psi}$ can be solved quite easily.
The two pairs of fermionic variables can be reduced to one pair. The symplectic structure (\ref{symplectic1}) is then finally reduced to:
\begin{eqnarray}
\int [E^a_i\partial_t (K_a^i-\alpha\widehat{\Gamma}_a^i) + \left( \partial_t {\zeta}^\dagger \right)
\Pi_{\zeta^\dagger} + \Pi_\zeta \left( \partial_t \zeta \right)]=\int[E^a_i\partial_t (-\alpha A^{i}_{a}) -2i\zeta^{\dag}\partial_{t}\zeta ], \label{symplectic3}
\end{eqnarray}
where we define $A^{i}_{a}:=\widehat{\Gamma}^i_a -\frac{1}{\alpha} K_a^i$ and have
also neglected the total time derivative terms. All the second-class constraints have now been solved and we have finally obtained the basic phase space variables on the "reduced phase space".
Note that $A_a^i$, the variable conjugate to $E^a_i$, is exactly same as the Ashtekar-Barbero connection obtained in standard analysis. We have not yet shown that it is a connection. We will do so in the next section.


\section{$SU(2)$ Gauge Theory}\label{sec5}

We have obtained a consistent Hamiltonian system which is invariant under local $SU(2)$ rotations. However this is not a $SU(2)$ gauge theory yet. The basic variables in the gravitational sector are the densitized triad $E^a_i$ and its conjugate $A_a^i$. The spin connection
$\Gamma_a^i$, given by Eq. (\ref{gammasol1}), is a function of $E^a_i$, $\zeta$ and $\zeta^\dagger$. In this section, we shall rewrite the remaining first class constraints in terms of the new variables.

First, let us consider the Gaussian constraint (\ref{gauss2}). In time gauge, using the constraint equation (\ref{newconstraint1}) and the solutions of $\Phi^a_{IJ}$ we can rewrite it as
\begin{eqnarray}
 \mathcal{G}_{0i} &=&\partial_a E^a_i + \epsilon_{ijk} \Gamma^j_a E^{a k}+  \frac{\alpha}{4} \left( \epsilon_{ijk}K^j_b E^{b k}- \zeta^\dagger \gamma^0 \gamma_5 \gamma_i
\zeta \right) \approx 0 ,\label{g0i} \\
\epsilon^{ijk}\mathcal{G}_{jk} &=&\alpha \left( \partial_a E^a_i + \epsilon_{ijk}
\Gamma^j_a E^{a k}\right)- \epsilon^{ijk}K_{aj} E^a_{k}+\zeta^\dagger \gamma^0 \gamma_5 \gamma_i \zeta \approx0. \label{gij}
\end{eqnarray}
Recall that the Gaussian constraint was used in getting Eq.(\ref{torspinreln}). Then in the gauge fixed theory, $\mathcal{G}_{0i}$ is explicitly resolved together with the second class constraint (\ref{newconstraint1}) by Eqs. (\ref{ca0sol}) and (\ref{caisol}). Hence, in terms of the new variables, it can be easily seen that $\mathcal{G}_{0i}$ is
identically zero as expected. Also comparing Eq. (\ref{gij}) with Eq. (\ref{ca0sol}) it is easy to see that $\epsilon_{jkl}
\mathcal{G}_{jk} \approx 0$ implies $C^a_0\approx 0$ (assuming $E^a_i \neq 0$).

Let us define
\begin{eqnarray}
G_{i}&:=&\frac{1}{\alpha}  \epsilon^{ijk}\mathcal{G}_{jk}  \nonumber\\
&=&\partial_a E^a_i +  \epsilon_{ijk}(\widehat{\Gamma}^j_a -\frac{1}{\alpha} K_a^j) E^{ak}+\frac{1}{\alpha}\zeta^\dagger \gamma^0 \gamma_5 \gamma_i \zeta\nonumber\\ &=&\partial_a E^a_i +  \epsilon_{ijk}A_a^j E^{ak}+\frac{1}{\alpha}\zeta^\dagger \gamma^0 \gamma_5 \gamma_i
\zeta \nonumber\\ &\equiv&\mathcal{D}_a E^a_i +\frac{1}{\alpha}\zeta^\dagger \gamma^0 \gamma_5 \gamma_i \zeta\approx 0, \label{newgauss}
\end{eqnarray}
Therefore $A_a^i \equiv \widehat{\Gamma}^i_a -\frac{1}{\alpha} K_a^i$ is the new connection, and using this connection we have obtained the Gaussian constraint in the standard $SU(2)$ gauge theory form. Tensorially, the new connection $A_a^i$ which we have defined is in the
same form as the standard Ashtekar-Barbero connection without torsion. The coupling parameter $\alpha$ plays the role of the Barbero-Immirzi parameter of the standard treatment. Hence the Barbero-Immirzi parameter in loop quantum gravity
acquires its physical meaning as the coupling constant between the Hilbert-Palatini term and the quadratic torsion term through our formulation.

Let us briefly recap what we have done. The basic variable $\Pi^a_I$ which encoded the torsion has been solved in terms of the fermionic degrees of
freedom using the constraints $\mathcal{G}_{IJ},C^a_I,\Phi^a_{IJ},\chi^{ab}$. We had started with a $SO(1,3)$
connection $\omega_a^{~IJ}$ which is not torsion free. That fact is reflected in our expression of the $SU(2)$ spin connection $\Gamma_a^i$ in Eq. (\ref{gammasol2}). But in the new connection $A_a^i$ which we define above, we remove exactly that additional piece.
However, since we had defined $K_a^i := 2 \omega_a^{~0i}$, the variable $K_a^i$ implicitly contains information about the torsion, though this is not obvious from Hamiltonian formulation itself. When there is no matter, torsion goes to zero and the $S_T$ term in our action (\ref{action}), and therefore, the terms originating from it
in the Hamiltonian analysis vanish \cite{1stpaper}. Then we go back to the standard formalism with a torsion-free $SO(1,3)$ spin connection.

We have obtained a $SU(2)$ gauge theory formulation of our system. The remaining constraints $H_a~,~H$ can also be rewritten in terms of the new basic variables. Using $K^{i}_{a}=-\alpha(A^{i}_{a}-\widehat{\Gamma}_a^i)$ and the Gaussian constraint (\ref{newgauss}), the
vector constraint (\ref{diffeo1}) can be written as
\begin{eqnarray}
 H_{a} &=& E^b_i\partial_{[a} K_{b]}^i-K^i_a\partial_b E_i^b-i\bigg(\zeta^{\dag}\partial_{a}\zeta-(\partial_{a}\zeta^{\dag})\zeta \bigg) \nonumber \\ &\approx& -\alpha E^{b}_{i}F^{i}_{ab}-
A_a^i \zeta^\dagger \gamma^0 \gamma_5 \gamma_i \zeta -i\bigg(\zeta^{\dag}\partial_{a}\zeta-(\partial_{a}\zeta^{\dag})\zeta \bigg), \label{newdiffeo}
\end{eqnarray}
 where $F^{i}_{ab}:= \partial _{[a}A^{i}_{b]}+\epsilon^{i}_{jk}A^{j}_{a}A^{k}_{b}$ is the curvature of
$A^{i}_{a}$. This is exactly the standard form of the vector constraint.  The Hamiltonian constraint (\ref{hamiltonian1}) is more complicated. After some calculation, we get
\begin{eqnarray}
 H&=& \frac{1}{\sqrt{q}}\left[\epsilon_{ijk}E^{a}_{i}E^{b}_{j}
F^{k}_{ab}-\left(\frac{1}{\alpha^{2}}+\frac{1}{4}\right)E^{a}_{i}E^{b}_{j}K^{i}_{[a}K^{j}_{b]}\right] +\frac{1}{2\sqrt{q}}\zeta^{\dag}\gamma_{5}\zeta \epsilon_{ijk}E^{a}_{i}E^{b}_{j}\partial_{a} E^{k}_{b} \nonumber\\ & &
+\frac{9}{8\sqrt{q}}\bigg(\zeta^{\dag}\gamma_{5}\zeta\bigg) \bigg(\zeta^{\dag}\gamma_{5}\zeta\bigg) +\frac{2}{\sqrt{q}}\bigg(\zeta^{\dag}\tau_{i}\zeta\bigg) \bigg(\zeta^{\dag}\tau_{i}\zeta\bigg) \nonumber\\ & &
+\frac{4i}{\alpha}\partial_{a}\left(\frac{1}{\sqrt{q}}E^{a}_{i}\zeta^{\dag}\tau_{i}\zeta\right)+ \frac{2iE^{a}_{i}}{\sqrt{q}}\bigg((\partial_{a} \zeta^{\dag})\sigma^{0i}\zeta-\zeta^{\dag}\sigma^{0i}\partial_{a}\zeta \bigg). \label{newham}
\end{eqnarray}
where
$\tau_{i}=-\frac{i}{2}\sigma_{i}$ with $\sigma_i$ being Pauli matrices. This expression goes over to the standard expression when the fermions are set to zero. Thus we complete our task of obtaining a $SU(2)$ gauge theory. Note that, since we have not split the
connection into torsion dependent and torsion free parts, it is not obvious how to directly compare the expressions of our constraints with those obtained in Ref.\cite{martinfermion}.


\section{Conclusion}\label{sec6}

Let us briefly summarize what we have achieved in this paper. We started with the action (\ref{action}) containing a torsion-squared term and fermionic matter apart from the standard Hilbert-Palatini term. This $T^2$ term is just the difference between the total
derivative Nieh-Yan term and the Holst term. Since an $SU(2)$ gauge theory formulation can be derived from actions containing either \cite{holst,shyam}, it seemed possible that such a formulation can also be obtained from our action containing only the $T^2$ term. We
also need to add fermionic matter because the vacuum case is torsion free \cite{1stpaper} and we are left with only the well-known Hilbert-Palatini part. We take non-minimally coupled fermionic matter so that the classical equations of motion for the fermions may not
depend on the coupling constant $\alpha$ multiplying the torsion term under certain condition. The equations of motion of the coupled system are derived. It is confirmed that, under the ansatz $\varepsilon=\frac{\alpha}{2}$, the dynamical system we obtain is equivalent to the standard Palatini formulation of GR minimally coupled to fermions.

We do a $3+1$ decomposition of our action, do a constraint analysis with the above ansatz and finally obtain a consistent Hamiltonian system with second-class constraints. All the second-class constraints are solved after breaking the $SO(1,3)$ invariance by fixing time gauge. As far as we
know, such Hamiltonian analysis on an action with non-zero torsion term with explicit expressions of all the second-class constraints is new in literature. Similar analysis with the Holst term (with non-minimally coupled fermions)
\cite{perezrovelli,mercuri,martinfermion} and the Nieh-Yan term (with minimally coupled fermions) \cite{shyam} has already been attempted before. Apart from the crucial fact that the gravitational part of our action is different
from those studied in literature so far, there are several other differences in our approach. Since we are motivated by PGT where the initial action is invariant under local Poincare transformations, our starting variables are different from those used in
Refs.\cite{perezrovelli,mercuri}. Unlike the treatment in Ref.\cite{martinfermion} we do not break up our variables into the torsion dependent and independent pieces. Moreover, since we do not have the Holst term, the techniques developed in Ref.\cite{peldan} for dealing
with second-class constraints and used in Refs.\cite{shyam,martinfermion} are not available to us. Also, unlike the treatments in Refs.\cite{perezrovelli,mercuri,martinfermion}, we fix the time gauge after we have found all the second-class constraints and obtained a consistent
Hamiltonian system. Furthermore, the Barbero-Immirzi parameter in loop quantum gravity obtains a new understanding in our formulation.

On solving the basic variable $\Pi^{a}_{I}$, torsion gets related to the fermionic degrees of freedom via Eq.(\ref{torspinreln}) which is as same as the one obtained in Ref.\cite{shyam}. Further, solution of the second-class constraint $C^a_i$ gives the $SU(2)$
spin connection $\Gamma_a^i$ in terms of the densitized triad. This differs from the spin connection in GR \cite{thiemannbook} only by a term which depends on the fermions. In the final step we obtain the connection dynamics by defining a new connection $A_a^i$ which is
algebraically in the same form as the Ashtekar-Barbero connection without torsion. However, unlike the torsion-free case, the $K_a^i$ part comes from the $\omega^{~0i}_a$ part of the $SO(1,3)$ connection which is not torsion free. As a result it is not obvious if
$K_a^i$ can be directly related to the extrinsic curvature $K_{ab}$ on shell. While the vector constraint (\ref{newdiffeo}) is standard, the additional terms in our Hamiltonian constraint (\ref{newham}) are somehow different from the ones obtained in literature.
Although these constraints can be loop quantized using existing techniques, it may be possible to rewrite them in a form more convenient for loop quantization. We leave this issue for future research. The present work at least opens the door to extending loop
quantization techniques from standard GR to more general PGT of gravity.

It should be remarked that, as a result of non-minimally coupled fermionic matter and the ansatz $\varepsilon = \frac{\alpha}{2}$, the coupling
constant $\alpha$ in our starting action totally disappears from the Lagrangian equations of motion. However, in the Hamiltonian connection formalism, the new connection $A_a^i$ and thus the constraints depend on the parameter $\alpha$ explicitly. This is somehow required by the $SU(2)$ gauge theory
formulation. A similar case happens also in the connection dynamics derived from the generalized Palatini action. It is still interesting to consider the general case when the two coupling parameters are not related to each other and thus the gauge theory is different from Palatini theory. We leave this open issue for future study. Nevertheless,
an enlightening result of our formulation is that the Barbero-Immirzi parameter in loop quantum gravity
acquires its physical meaning as the coupling constant between the Hilbert-Palatini term and the quadratic torsion term. In fact, this
parameter measures the relative contribution of torsion in comparison with curvature in the action for this Poincare gauge theory of gravity.
However, it should be noted that, the $SU(2)$ gauge theory which we obtain is based on the particular choice of basic canonical variables by Eq.(\ref{symplectic3}). An alternative choice is to cancel all the $\alpha$-dependent terms in Eq.(\ref{symplectic1}). Then it is still possible to obtain, via a canonical
transformation, a $SU(2)$ gauge theory in which the connection does not depend on the coupling parameters of the starting action but on an arbitrary
constant appearing in the canonical transformation.

\begin{acknowledgements}

This work is supported in part by  NSFC (Grant Nos. 10975017 and 11235003) and the Fundamental Research Funds for the Central Universities. JY would like to acknowledge the support of NSFC (Grant No. 10875018). KB would also like to thank China Postdoctoral Science
Foundation (Grant No.20100480223) as well as DST-Max Planck Partner Group of Dr. S. Shankaranarayanan, IISER, Thiruvananthapuram for financial support.

\end{acknowledgements}


\appendix \section{Gamma Matrix}\label{app1}

In this section we collect some of the standard properties of Dirac matrices which we have used in previous sections. The $\gamma$ matrices, in any dimension, satisfy the Clifford algebra
\begin{eqnarray}
\{ \gamma_I,\gamma_J \} = \gamma_I \gamma_J + \gamma_J \gamma_I
= 2 \eta_{IJ}
 \end{eqnarray}
 where $\eta_{IJ}$ is the flat Minkowski metric. We shall restrict ourselves to 4 dimensions and choose the signature $(-+++)$ which is different from the signature usually used in QFT. In this signature the above relation can be decomposed
as
\begin{eqnarray}
{\gamma_0}^2 = - {\mathbb I}_4 \hspace{2em};\hspace{2em} {\gamma_i}^2 =  {\mathbb I}_4.  \nonumber
\end{eqnarray}
 This implies that $\gamma_0$ is anti-Hermitian while $\gamma_i$ is Hermitian. Note that all the $\gamma$ matrices are unitary. We
also define the commutator $\sigma_{IJ} := \frac{1}{4}[\gamma_I,\gamma_J]$ and another standard combination $\gamma_5 := i \gamma_0 \gamma_1 \gamma_2 \gamma_3$. It is easy to check that $(\gamma_5)^2 = {\mathbb I}$ and $(\gamma_5)^\dag = \gamma_5$. In the Weyl
representation, commonly used for massless fermions, the Dirac matrices can be explicitly written as
\begin{eqnarray}
\gamma_0 = \left( \begin{array}{cc} 0 & i{\mathbb I}_2 \\ i{\mathbb I}_2 & 0 \end{array} \right) \hspace{2em};\hspace{2em} \gamma_i = \left(
\begin{array}{cc} 0 & -i\sigma_i \\ i \sigma_i & 0 \end{array} \right) \hspace{2em};\hspace{2em} \gamma_5 = \left( \begin{array}{cc}  - {\mathbb I}_2 & 0 \\ 0 &{\mathbb I}_2 \end{array} \right). \nonumber
\end{eqnarray}
In this paper we have used the following standard
identities
\begin{eqnarray}
\left\{ \gamma_5 , \gamma_I \right\} ~~= ~~&0& ~~=~~\left[ \gamma_5 ,\sigma_{IJ} \right] ; \nonumber\\
  \left[ \gamma_K,\sigma_{IJ} \right] = \eta_{K[I}\gamma_{J]}  \hspace{2em}&;&\hspace{2em}
\left\{ \gamma_K, \sigma_{IJ} \right\} = i \epsilon^{KIJL} \gamma_5 \gamma_L. \label{gammatprop1}
\end{eqnarray}


\section{$3+1$ Decomposition}\label{app2}

In this section we give the parametrization of the tetrad and the co-tetrad fields which we use in this paper. They read \cite{peldan}
\begin{eqnarray}
e^{t I} = - \frac{N^I}{N} ~~~~ &;& ~~~~  e^{a I} = V^{a I} + \frac{N^a N^I}{N},\nonumber \\ e_{t I} = N N_I + N^a
V_{a I} ~~~~ &;& ~~~~  e_{a I} = V_{a I},\label{newparameters1}\\ \mbox{with}\hspace{4em} N^I V_{a I} = 0 ~~~~ &;& ~~~~  N^I N_I =-1 . \label{parameterconstraints1}
\end{eqnarray}
What we have done is that we have reparametrized the 16 degrees of freedom of $e_{\mu I}$
into 20 fields given by (\ref{newparameters1}) subject to the 4 relations (\ref{parameterconstraints1}). Note that this is just a convenient reparametrization of the initial variables. From these definitions, the following identities also hold:
\begin{eqnarray}
 &&V^{a
I} V_{b I} = \delta^a_b ~~~~ ; ~~~~  V^{a I}N_I = 0  ~~~~ ; ~~~~ N_a := V_{a I} V_{b}^I N^b ,\nonumber\\ && V^{aI} V_{a}^{~J} = \eta^{IJ} + N^I N^J . \label{parameterconstraints2}
\end{eqnarray}
In terms of these fields the metric takes the standard form
\begin{eqnarray}
g_{\mu \nu} = \left(\begin{array}{cc}
              -N^2 + N^a N_a  &    N_a    \\
               N_a            &  V_{a I} V_b^I
           \end{array}\right)  .                 \nonumber
\end{eqnarray}
It is easy to see that
\begin{eqnarray}
 g &:=& \mbox{det} (g_{\mu \nu}) = - N^2 \mbox{det}( V_{a I} V_b^I ),\nonumber \\ e &:=& |\mbox{det} (e_{\mu I})| = N \sqrt{\mbox{det}( V_{a I} V_b^I) }  = N \sqrt{\mbox{det}( q_{ab})} = N \sqrt{q}. \nonumber
\end{eqnarray}
Using the definitions given above we can also prove the following two identities which have been used in our analysis,
\begin{eqnarray}
 - e e^{a}_{[I} e^{b}_{J]} &=& \frac{N^2}{e} \Pi^{[a}_{IK} \Pi^{b]}_{JL} \eta^{KL} + N^{[a}\Pi^{b]}_{IJ} ,
\label{identity1} \\ \Pi^a_{~IJ} &=&  \sqrt{q} V^a_{[I} N_{J]} ~ ~ \Rightarrow ~  ~V^a_I = - \frac{1}{\sqrt{q}}\Pi^a_{~IJ} N^J.  \label{identity2}
\end{eqnarray}


\section{Determination of $\lambda_a^{IJ}$} \label{app3}

In this section we show how to obtain $\lambda^{IJ}_a$ from Eqs. (\ref{Cdot}) and (\ref{chidot}). First let us consider $\dot{C}^a_I$. Using Eqs.(\ref{constalg2}-\ref{constalg5}), we get
\begin{eqnarray}
\dot{C}^a_I(\eta_a^I) &=& \left\{C^a_I (\eta_a^I), \left( \Phi^a_{IJ}( \lambda_a^{IJ}) + H(N)\right) \right\} \nonumber \\ &=& \int_{\Sigma}d^3x\epsilon^{abc} \eta_b^I V_c^J
\left[\left(\alpha \lambda_a^{IJ} +\epsilon_{IJKL}\lambda_a^{KL}\right) - \frac{\alpha N}{\sqrt{q}} \left( \overline{D_a \lambda} ~\sigma_{IJ} \Pi -\overline{\Pi} ~\sigma_{IJ} D_a \lambda \right) \right] ~ ~  \hspace{1em} \forall ~ \eta_b^I. \label{cdotproof1}
\end{eqnarray}
For convenience, we define
\begin{eqnarray}
 X_a^{IJ} := \left(\alpha \lambda_a^{IJ} +\epsilon_{IJKL}\lambda_a^{KL}\right) - \frac{\alpha N}{\sqrt{q}} \left( \overline{D_a \lambda} ~\sigma_{IJ} \Pi -\overline{\Pi} ~\sigma_{IJ} D_a \lambda \right).
\label{XaIJ1}
\end{eqnarray}
Thanks to this definition, it is easy to see from Eq.(\ref{cdotproof1}) that
\begin{eqnarray}
\dot{C}^a_I ~ \approx 0 ~ \Rightarrow ~  \epsilon^{abc}  V_c^J X_a^{IJ} = 0.  \label{cdotproof2}
\end{eqnarray}
Eq.(\ref{XaIJ1}) can easily be inverted to express $\lambda_a^{IJ}$ in terms of $X_a^{IJ}$ as
\begin{eqnarray}
\lambda_a^{IJ} = \frac{\alpha}{\alpha^2 + 4}\left[ \frac{\alpha N}{\sqrt{q}}
\left( \overline{D_a \lambda} ~\sigma_{IJ} \Pi -\overline{\Pi} ~\sigma_{IJ} D_a \lambda \right)  - \frac{N}{\sqrt{q}}\epsilon_{IJKL} \left( \overline{D_a \lambda} ~\sigma^{KL} \Pi -\overline{\Pi} ~\sigma^{KL} D_a \lambda \right) + X_a^{IJ} - \frac{1}{\alpha}
\epsilon_{IJKL} X_a^{KL} \right]. \nonumber\\ \label{XaIJ2}
\end{eqnarray}
Now let us consider $\dot{\chi}^{ab}$. We have
\begin{eqnarray} \dot{\chi}^{ab}(\sigma_{ab})  &=& \left\{\chi^{ab}(\sigma_{ab}),\Phi^a_{IJ}( \lambda_a^{IJ})\right\} + \left\{\chi^{ab}(\sigma_{ab}), \left(C^a_I (\gamma_a^I) +
H(N) \right)  \right\} \nonumber\\ &=&  -\int_{\Sigma}d^3x2 \sigma_{ab} \lambda_c^{IJ} \epsilon^{acd} \epsilon_{IJKL} V^b_K V_d^L +  \int_{\Sigma}d^3x\sigma_{ab} \Sigma^{ab}  ~ ~, \hspace{1em} \forall ~  \sigma_{ab}, \label{chidotproof1}
\end{eqnarray}
 where we have defined $\int_{\Sigma}d^3x\sigma_{ab} \Sigma^{ab} :=
\left\{\chi^{ab}(\sigma_{ab}), \left(C^a_I (\gamma_a^I) + H(N) \right)  \right\}$.
The explicit form of $\Sigma^{ab}$ is very complicated and can be calculated using Eqs.
(\ref{gammasoln2}), (\ref{constalg9}) and (\ref{constalg11}). However we do not need the explicit form of $\Sigma^{ab}$. We are interested in solving for $\lambda_a^{IJ}$ which only comes from the first part of Eq. (\ref{chidotproof1}). After some more algebra we obtain
the equation in terms of $X_a^{IJ}$ as
\begin{eqnarray} \dot{\chi}^{cd}  ~ \approx ~ 0 ~  \Rightarrow ~ &&
 \Sigma^{cd} + \frac{\alpha}{\alpha^2 +4} \bigg[ 2 V^c_I V^d_I \Pi^a_{KL} X_a^{KL} -
 \left( V^a_I V^d_I \Pi^c_{KL} + V^a_I V^c_I \Pi^d_{KL} \right) X_a^{KL}
\nonumber \\ && -\epsilon^{cab}\left(\frac{\alpha N}{\sqrt{q}} \epsilon_{IJKL} \left( \overline{D_a \lambda} ~\sigma^{KL} \Pi -\overline{\Pi} ~\sigma^{KL} D_a \lambda \right)  + \frac{4N}{\sqrt{q}}\left( \overline{D_a \lambda} ~\sigma^{IJ} \Pi -\overline{\Pi}
~\sigma^{IJ} D_a \lambda \right)\right)V^d_I V_b^J \nonumber \\ && -\epsilon^{dab}\left(\frac{\alpha N}{\sqrt{q}} \epsilon_{IJKL} \left( \overline{D_a \lambda} ~\sigma^{KL} \Pi -\overline{\Pi} ~\sigma^{KL} D_a \lambda \right)  + \frac{4N}{\sqrt{q}}\left( \overline{D_a
\lambda} ~\sigma^{IJ} \Pi -\overline{\Pi} ~\sigma^{IJ} D_a \lambda \right)\right) V^c_I V_b^J\bigg] \nonumber\\ && ~ \approx ~ 0.
 \label{chidotproof2}
\end{eqnarray}
Using Eqs.(\ref{cdotproof2}) and  (\ref{chidotproof2}), after a long calculation we get
\begin{eqnarray}
 X_a^{IJ} &=& \frac{1}{4 \sqrt{q}} \bigg[ V_{a[I}N_{J]} \epsilon^{def} A_d^{KL} V_{e K} V_{f L} + V_{c[I}N_{J]} \epsilon^{cef} A_e^{KL} V_{a K} V_{f L}
+  A_{e K[J } N_{I]} \epsilon^{def} V_{d L} V_{a}^L V_{f K} \bigg] \nonumber \\ && \hspace{4em} +~ \frac{\alpha^2 +4}{4 \alpha \sqrt{q}} \bigg[ \frac{1}{2} V_{a[I}N_{J]} \Sigma^{cd} V_c^K V_{d K} -  V_{c[I}N_{J]} \Sigma^{cd} V_a^K V_{d K} \bigg] \label{XaIJ3}
\end{eqnarray}
where, for brevity of notation, we have defined
\begin{eqnarray} A_a^{IJ} := \frac{\alpha N}{\sqrt{q}} \epsilon_{IJKL}\left( \overline{D_a \lambda} ~\sigma^{KL} \Pi -\overline{\Pi} ~\sigma^{KL} D_a \lambda \right)  + \frac{4N}{\sqrt{q}}\left(
\overline{D_a \lambda} ~\sigma^{IJ} \Pi -\overline{\Pi} ~\sigma^{IJ} D_a \lambda \right). \nonumber
\end{eqnarray}
Putting Eq.(\ref{XaIJ3}) into Eq.(\ref{XaIJ2}), we get $\lambda_a^{IJ}$.


\end{document}